\documentclass[11pt]{article}
\usepackage{xifthen}

    \renewcommand{\phi}{\varphi}
        \def\set#1{\left\{ #1 \right\}}

        \newcommand\PPF{{\mathbb P _f}}
        \newcommand\PPC{\ensuremath{\mathbb P _c}}

    \newenvironment{proof-sketch}{\medskip\noindent{\em Sketch of Proof.}\hspace*{1em}}{\qed\bigskip}
        \newenvironment{proof-attempt}{\medskip\noindent{\em Proof attempt.}\hspace*{1em}}{\bigskip}

        {\vspace{1ex}\noindent{\emph{Proof.}}\hspace{0.5em}\def\myproof@name{#1}\newif\ifqedhere\qedherefalse}%
        {\ifqedhere \else \hfill{\tiny \qed\ (\myproof@name)}\vspace{1ex} \qedherefalse \fi}

\usepackage{lipsum}
\usepackage{setspace}
\usepackage{mdframed}
\usepackage[usenames,pdf]{pstricks}
\usepackage{epsfig}
\usepackage{color}
\usepackage{amsmath}
\usepackage[normalem]{ulem}

        \usepackage[hyphens]{url}
        \usepackage[margin=1.0in]{geometry}
        \usepackage[noadjust]{cite}
        \usepackage[normalem]{ulem}
        \usepackage[small]{complexity}
        \usepackage{algpseudocode}
        \usepackage{amsfonts, amsmath, amssymb, amsthm}
        \usepackage{bbm}
        \usepackage{etoolbox}
        \usepackage{graphicx}
        \usepackage{intcalc}
        \usepackage{lmodern}
        \usepackage{microtype}
        \usepackage{nag}
        \usepackage{nicefrac}
        \usepackage[super]{nth}
        \usepackage{stmaryrd}
        \usepackage{tikz}
        \usepackage{tocbibind}
        \usepackage{version}
        \usepackage{xparse}
        \usepackage{xspace}
        \usepackage{xstring} 

        \usepackage[pdfstartview=FitH,pdfpagemode=None,colorlinks=true]{hyperref}
        \hypersetup{
                linkcolor=[rgb]{0.5, 0.2, 0.2},
                citecolor=[rgb]{0.2, 0.5, 0.2},
                urlcolor =[rgb]{0.2, 0.2, 0.5}
        }
        \usepackage{algorithm}

        \makeatletter
        \global\let\tikz@ensure@dollar@catcode=\relax
        \makeatother

        \SetSymbolFont{stmry}{bold}{U}{stmry}{m}{n}

        \PackageWarning{miforbes}{overfullrule enabled}
        \newcommand{\ignore}[1]{}
        \newcommand{\demph}[1]{\textbf{#1}}

        \newcommand*{\subproofname}{Sub-Proof:}

        \newcommand{\bits}{\ensuremath{\{0,1\}}}

        \DeclareMathOperator{\tr}{tr}

        \newlang{\PIT}{PIT}

        \PackageWarning{miforbes}{get rid of matrix commands}

        \renewcommand{\vec}[1]{\overline{#1}}

        \newcommand{\F}{\mathbb{F}}

        \newcommand{\cC}{\mathcal{C}}

        \makeatletter
        \newcommand{\va}{{\vec{a}}\@ifnextchar{^}{\!\:}{}}
        \newcommand{\vb}{{\vec{b}}\@ifnextchar{^}{\!\:}{}}
        \newcommand{\vc}{{\vec{c}}\@ifnextchar{^}{\!\:}{}}
        \newcommand{\vd}{{\vec{d}}\@ifnextchar{^}{\!\:}{}}
        \newcommand{\ve}{{\vec{e}}\@ifnextchar{^}{\!\:}{}}
        
        \newcommand{\vg}{{\vec{g}}\@ifnextchar{^}{\!\:}{}}
        \newcommand{\vh}{{\vec{h}}\@ifnextchar{^}{\!\:}{}}
        \newcommand{\vi}{{\vec{i}}\@ifnextchar{^}{\!\:}{}}
        \newcommand{\vj}{{\vec{j}}\@ifnextchar{^}{\!\:}{}}
        \newcommand{\vk}{{\vec{k}}\@ifnextchar{^}{\!\:}{}}
        \newcommand{\vl}{{\vec{\ell}}\@ifnextchar{^}{\!\:}{}}
        \newcommand{\vm}{{\vec{m}}\@ifnextchar{^}{\!\:}{}}
        \newcommand{\vn}{{\vec{n}}\@ifnextchar{^}{\!\:}{}}
        \newcommand{\vo}{{\vec{o}}\@ifnextchar{^}{\!\:}{}}
        \newcommand{\vp}{{\vec{p}}\@ifnextchar{^}{\!\:}{}}
        \newcommand{\vq}{{\vec{q}}\@ifnextchar{^}{\!\:}{}}
        \newcommand{\vr}{{\vec{r}}\@ifnextchar{^}{\!\:}{}}
        \newcommand{\vs}{{\vec{s}}\@ifnextchar{^}{\!\:}{}}
        \newcommand{\vt}{{\vec{t}}\@ifnextchar{^}{\!\:}{}}
        \newcommand{\vu}{{\vec{u}}\@ifnextchar{^}{\!\:}{}}
        \newcommand{\vv}{{\vec{v}}\@ifnextchar{^}{\!\:}{}}
        \newcommand{\vw}{{\vec{w}}\@ifnextchar{^}{\!\:}{}}
        \newcommand{\vy}{{\vec{y}}\@ifnextchar{^}{\!\:}{}}
        \newcommand{\vx}{{\vec{x}}\@ifnextchar{^}{}{}}          
        \newcommand{\vz}{{\vec{z}}\@ifnextchar{^}{\!\:}{}}

        \newcommand{\vA}{{\vec{A}}\@ifnextchar{^}{\!\:}{}}
        \newcommand{\vB}{{\vec{B}}\@ifnextchar{^}{\!\:}{}}
        \newcommand{\vC}{{\vec{C}}\@ifnextchar{^}{\!\:}{}}
        \newcommand{\vD}{{\vec{D}}\@ifnextchar{^}{\!\:}{}}
        \newcommand{\vE}{{\vec{E}}\@ifnextchar{^}{\!\:}{}}
        \newcommand{\vF}{{\vec{F}}\@ifnextchar{^}{\!\:}{}}
        \newcommand{\vG}{{\vec{G}}\@ifnextchar{^}{\!\:}{}}
        \newcommand{\vH}{{\vec{H}}\@ifnextchar{^}{\!\:}{}}
        \newcommand{\vI}{{\vec{I}}\@ifnextchar{^}{\!\:}{}}
        \newcommand{\vJ}{{\vec{J}}\@ifnextchar{^}{\!\:}{}}
        \newcommand{\vK}{{\vec{K}}\@ifnextchar{^}{\!\:}{}}
        \newcommand{\vL}{{\vec{L}}\@ifnextchar{^}{\!\:}{}}
        \newcommand{\vM}{{\vec{M}}\@ifnextchar{^}{\!\:}{}}
        \newcommand{\vN}{{\vec{N}}\@ifnextchar{^}{\!\:}{}}
        \newcommand{\vO}{{\vec{O}}\@ifnextchar{^}{\!\:}{}}
        \newcommand{\vP}{{\vec{P}}\@ifnextchar{^}{\!\:}{}}
        \newcommand{\vQ}{{\vec{Q}}\@ifnextchar{^}{\!\:}{}}
        \newcommand{\vR}{{\vec{R}}\@ifnextchar{^}{\!\:}{}}
        \newcommand{\vS}{{\vec{S}}\@ifnextchar{^}{\!\:}{}}
        \newcommand{\vT}{{\vec{T}}\@ifnextchar{^}{\!\:}{}}
        \newcommand{\vU}{{\vec{U}}\@ifnextchar{^}{\!\:}{}}
        \newcommand{\vV}{{\vec{V}}\@ifnextchar{^}{\!\:}{}}
        \newcommand{\vW}{{\vec{W}}\@ifnextchar{^}{\!\:}{}}
        \newcommand{\vY}{{\vec{Y}}\@ifnextchar{^}{\!\:}{}}
        \newcommand{\vX}{{\vec{X}}\@ifnextchar{^}{}{}}          
        \newcommand{\vZ}{{\vec{Z}}\@ifnextchar{^}{\!\:}{}}
        \makeatother

        \newcommand{\vnz}{{\vec{0}}}

\hypersetup{
        linkcolor=[rgb]{0.2,0.2,0.5},
        citecolor=[rgb]{0.2, 0.45, 0.2},
        urlcolor=[rgb]{0.6, 0.2, 0.2}
}

\usepackage{nth}
\usepackage{intcalc}
\usepackage{etoolbox}
\usepackage{xstring}

\newcommand{\ECCCurl}[2]{http://eccc.hpi-web.de/report/\ifnumcomp{#1}{>}{93}{19}{20}#1/#2/}

\newcommand{\shortECCC}[2]{\texttt{\href{http://eccc.hpi-web.de/report/\ifnumcomp{#1}{>}{93}{19}{20}#1/#2/}{eccc:TR#1-#2}}}

\newcommand{\parseECCC}[1]{
\StrSubstitute{#1}{TR}{}[\tmpstring]%
\IfSubStr{\tmpstring}{/}{ 
\StrBefore{\tmpstring}{/}[\ecccyear]%
\StrBehind{\tmpstring}{/}[\ecccreport]%
}{
\StrBefore{\tmpstring}{-}[\ecccyear]%
\StrBehind{\tmpstring}{-}[\ecccreport]%
}%
\shortECCC{\ecccyear}{\ecccreport}}

\PackageWarning{miforbes}{LATER: is deferred}

\newcommand{\LATER}[1]{}

        \newcommand{\aaaa}[1]{}
        \newcommand{\Pudlak}{Pudl{\'{a}}k\xspace}
        \newcommand{\Krajicek}{Kraj\'{i}\v{c}ek\xspace}
        \makeatletter   
        \renewcommand\paragraph{\@startsection{paragraph}{4}{\z@}%
                {1ex \@plus1ex \@minus.2ex}%
                {-1em}%
                {\normalfont\normalsize\bfseries}}
        \makeatother

        \newtheorem{theorem}{Theorem}[section]
        \newtheorem*{theorem*}{Theorem}
        
        \newtheorem*{theoremwp*}{Theorem}
        \newtheorem{lemma}[theorem]{Lemma}

        \newtheorem{definition}{Definition}[section]

\def\_{\,\,\,\,\,}

\def\zo{ \{0,1\} }

\allowdisplaybreaks

\newcommand{\cd}{\cdot}

\newcommand{\mar}[1]{}

\newcommand{\para}[1] {\paragraph{#1}}

\def\RCD0#1{{\mbox{\rm R$_{#1}$(lin)}}}

\newcommand{\ACZ}{{\rm AC}\ensuremath{^0}}

\newcommand{\QuadSpace}{\smallskip}
\newcommand{\HalfSpace}{\medskip}

\newcommand{\lIPS}{\texorpdfstring{IPS$_{\text{LIN}}$}{IPS-LIN}\xspace}

\renewcommand{\cc}[1]{\ensuremath{\mathsf{#1}}} 

\newcommand{\definedWord}[1]{\emph{#1}}

\author{%
        Toniann Pitassi~\thanks{Email: \texttt{toni@cs.toronto.edu}. Department of Computer Science, University of Toronto, Toronto, Canada.}
        \and 
        Iddo Tzameret~\thanks{Email: \texttt{iddo.tzameret@rhul.ac.uk}. Department of Computer Science, Royal Holloway, University of London, Egham, UK.}
}
\date{}

\begin{document}

\title{Algebraic Proof Complexity: Progress, Frontiers and Challenges\footnote{A version of this paper appears in the Complexity Column of the \textit{ACM SIGLOG News}, ACM New York, NY, USA, July 2016.}}
\maketitle

\begin{abstract}
We survey recent progress in the  proof complexity of strong proof systems and its connection
to algebraic circuit complexity, showing how the synergy between the two gives rise to new approaches to fundamental open questions, solutions to old problems, and  new directions of research. In particular, we focus on tight connections between proof complexity lower bounds (namely, lower bounds on the size of proofs of certain tautologies), algebraic circuit lower bounds, and the Polynomial Identity Testing problem from  derandomization theory. 
\end{abstract}
\small
\tableofcontents
\normalsize

\section{Introduction}
Propositional proof complexity aims to understand and analyze the computational resources required to prove propositional tautologies, in the same way that circuit complexity studies the resources required to compute boolean functions. 
A central question in the area asks whether every boolean tautology has a short propositional proof. 
Here, a propositional proof system can take many forms. One such proof system is the resolution refutation system whose proof-search algorithm constitutes the basis of current state of the art industrial-level SAT solvers (this thread of research was recently covered in Nordstr\"{o}m \cite{Nor15-siglog}). For resolution and its \textit{weak }extensions, strong lower bounds are known since Haken \cite{Hak85}. But the major open questions in proof complexity, those originating from boolean circuit complexity and complexity class separations, such as \P\ vs.~\NP, are about the length of \textit{much stronger} proof systems than resolution, and these stronger systems will be the focus of this survey.  

The prototypical strong proof system is the standard Hilbert-style propositional proof system, called \emph{Frege proof system}, in which a proof starts from a fixed finite set of axioms and derives new propositional \textit{formulas}
using a fixed set of sound derivation rules. 
Establishing  any super-polynomial size lower bound on such proofs (in terms of the size of the formula proved) is a major open problem in proof complexity, and  a  fundamental question in complexity theory. 


The seminal work of Cook and Reckhow~\cite{CookReckhow79} showed that in its strongest form, proof-size
lower bound questions relate  directly to  fundamental hardness questions in computational complexity: establishing super-polynomial lower bounds for \emph{every} propositional proof system would separate \NP\ from \coNP\ (and thus also \P\ from \NP). 

The aim of this survey is to outline a  new research direction, connecting algebraic circuit complexity to proof complexity. The prominent goal of this approach is the quest for lower bounds on strong proof systems; other important aspects are connections to derandomization theory and application to feasible mathematics. The survey is meant to give some basic background in propositional
proof complexity and describe the algebraic approach to proof complexity.
%

In what follows, Section \ref{sec:basics} gives the basic definitions of algebraic
circuits, propositional proof systems
and algebraic proof systems, and a quick survey of the background results. Section \ref{sec:IPS} is devoted to the Ideal Proof System (IPS). In this section we show that IPS is closely connected to the Extended Frege proof system, and show that superpolynomial lower bounds for IPS proofs imply
algebraic circuit lower bounds. 
Section \ref{sec:noncommIPS} is devoted to the study of the non-commutative IPS. In this section we show that non-commutative IPS is equivalent (up to quasi-polynomial
factors) to the Frege system
and discuss the ramifications of this equivalence.
Section \ref{sec:lower bounds}
is dedicated to lower bounds for restricted subsystems of IPS. Section \ref{sec:PIT} discusses connections between algebraic proof complexity and the polynomial identity testing (PIT) problem and the use of structural results on algebraic circuits in proof complexity, as well as application to feasible mathematics. Finally we conclude with a discussion and open problems in Section \ref{sec:conclusion}.


\section{Basic Concepts}\label{sec:basics}

For a natural number we let $[n]=\set{1,\ldots,n}$.
Let $ \F $ be a field. Denote by $ \F[x_1,\ldots,x_n] $ the ring of (commutative) polynomials with coefficients from $ \F $ and variables $ x_1,\ldots,x_n $. In this survey, unless
otherwise stated, we treat polynomials as \emph{formal }linear combination of monomials, where a monomial is a product of variables. Hence, when we talk about the \textit{zero polynomial} we mean the polynomial in which the coefficients of all monomials are zero  (it can happen that over, say, $GF(2)$, $x^2+x$ computes the zero \emph{function}, but it is \emph{not} the zero polynomial, because it has two nonzero monomial coefficients). Similarly, two polynomials are said to be \emph{identical} if they have precisely the same monomial coefficients. The \emph{degree} of a polynomial (or total degree) is the maximal sum of variable powers in a monomial with a nonzero coefficient in the polynomial. If the power
of each variable in every monomial is at most 1 we say that the polynomial is \emph{multilinear}. We write $\poly(n)$ to denote a polynomial growth in $n$, namely a function that is upper bounded by $n^{O(1)}$, and $\qpoly(n)$ to denote a quasi-polynomial growth in $n$, that is, $n^{\log^{O(1)}{\!n}}$.

%

\subsection{Propositional Proof Systems}
\label{sec:proof-complexity-background}

Cook and Reckhow \cite{CookReckhow79} defined a general
concept of a propositional proof system from the perspective
of computational complexity theory: a propositional
proof system is a polynomial-time function $f$ from a set of finite strings over some given alphabet \emph{onto} the set of propositional tautologies (reasonably encoded). Thus, $f(x)=y$ means that the string $x$ is a proof of the tautology $y$. Note that since \(f\) is onto, all tautologies and only tautologies have proofs (and thus the proof system is complete and sound). 

The idea behind the Cook-Reckhow definition is that a purported proof $x$ may be much longer than the tautology $y$ it proves, but given a proof it should be possible to efficiently check (efficient with respect to the \textit{proof length}) that it is indeed a correct proof of the tautology. We say that a propositional proof system
is \textit{polynomially bounded} if there exists a
polynomial $p$ that bounds the minimal proof size $|x|$ for every tautology $y$; namely, for every tautology $y$
its minimal proof $x$ is such that $|x|\le \poly(|y|)$.
Under the general Cook-Reckhow definition we have:

\begin{theorem}[Cook-Reckhow \cite{CookReckhow79}]\label{thm:Cook-Reckhow}
\NP$=$\coNP\ if and only if there is a polynomially bounded propositional proof system.
\end{theorem}

Therefore, proving lower bounds against stronger and
stronger propositional proof systems is clearly a formidable problem, as it can be considered as  partial progress towards proving
 \NP$\neq$\coNP\ (and thus \P$\neq$\NP). 
 
The definition of a propositional Cook-Reckhow proof system  encompasses most standard proof systems
for propositional tautologies, such as resolution and usual textbook proof system for propositional
logic. In this survey 
we discuss specific propositional proof
systems,  that are at least as strong as the Frege or the Extended Frege system (see below). 
Though proving lower bounds on (Extended) Frege proof sizes for some families of tautologies would not amount to \NP$\neq$\coNP,
it would still constitute a breakthrough in complexity theory. 


\subsubsection{Frege Proof Systems}
One of the most investigated and central propositional proof systems comes from the tradition of logic and is called the  \textit{Frege proof system}. 
A Frege proof system is any system that has a fixed number of axiom schemes and sound derivation rules, that is also implicationally complete\footnote{Meaning that if a set of formulas $\Gamma$ logically implies a formula $\varphi$, then there is a proof of $\varphi$ in the system with formulas in $\Gamma$ added to the axioms.}, %
and in which proof lines are written as propositional \textit{formulas}.
It is known since Reckhow's work \cite{Reckhow76} that all Frege proof systems 
are polynomially equivalent to each other, and hence it does not matter precisely which rules, axioms, and logical-connectives we use in the system.
For concreteness, the reader can think of \emph{the} \textit{\textbf{Frege} \textbf{proof system} }as the following simple one (known as \textit{Schoenfield's system}),  consisting of only three axiom schemes (where $A\to B$ is an abbreviation of $\neg A\lor B$; and $A,B,C$ are
any propositional formulas): \vspace{-5pt} 
\begin{gather*}
A\to(B\to A)
\\
(\neg A \to \neg B)\to ((\neg A\to B)\to A)
\\
(A\to (B\to C))\to ((A\to B)\to (A\to C)),
\end{gather*}
and a single inference rule (known as \textit{modus ponens}):
$$
\hbox{from $A$ and $A\to B$, infer $B$}\,.
$$

Frege systems are considered strong for several reasons. 
First, no super-polynomial lower bounds are known for Frege proofs, and moreover proving such lower bounds
seems to be out of reach of current techniques, and believed by some to be even harder
than proving explicit circuit lower bounds \cite{Razb15-annals}.
Secondly, hard candidates for Frege systems are hard to find; common tautologies such as
the pigeonhole principle that are known to be hard for weaker proof systems have
polynomial-size Frege proofs.
(See \cite{BBP95,Razb15-annals,Kra:book11,LT13} for further discussions on hard proof complexity candidates.)
%
%


An \textbf{\textit{Extended Frege}} proof system is obtained by augmenting
Frege with the axiom: 
\[
  \hbox{Extension Axiom:~~~~~~~~~}   z \leftrightarrow \phi\;,\
\]
where $z$ is any \textit{new} variable (namely, a variable that does
not occur before in the proof) and $\phi$ is any formula
(that does not contain $z$), and where the new variables $z$ appearing in the extension axiom does not occur in the final formula in the proof. 
The point of the extension axiom is to allow the use of
new variables to represent intermediate subformulas in a proof;
with this new axiom scheme, polynomial-size Extended Frege proofs can reason
about propositions computable by polynomial-size circuits
(rather than just propositions computable by polynomial-size formulas, as is
the case for polynomial-size Frege proofs).

For comprehensive texts on proof complexity and strong
proof systems see e.g., the
monograph by Kraj\'{i}\v{c}ek \cite{Kra95} and \cite[Chapter
5]{CK02}.


\subsection{Comparing Proof Systems}


To compare the relative strength of two proof systems we define the notion of a simulation.
We say that a propositional proof system $P$ \demph{polynomially simulates} another propositional proof system $Q$ if there is a polynomial-time computable function $f$ that maps $Q$-proofs to $P$-proofs of the same tautologies (if $P$ and $Q$ use different representations for tautologies, we fix a (polynomial) translation  from one representation to the other).  In case $f$ is computable in time $t(n)$ (for $n$ the input-size), we say that $P$ \emph{~$t(n)$-simulates} $Q$. We say that $P$ and $Q$ are \emph{polynomially equivalent} in case $P$ polynomially simulates $Q$ and $Q$ polynomially simulates $P$. If $P$ polynomially simulates $Q$ but $Q$ does not polynomially simulate $P$ we say that \emph{$P$ is strictly stronger than $Q$} (equivalently, that $Q$ \emph{is strictly weaker than $P$}).


\subsection{Algebraic Circuits, Formulas, and Algebraic Complexity Classes}\label{sec:algebraic_circuits} 
Algebraic circuits and formulas (over some  fixed chosen field or ring) compute polynomials via addition and multiplication gates, starting from the input variables and constants from the field. More precisely, an \emph{algebraic circuit} $F$ is a finite directed acyclic graph with \textit{input nodes}  (i.e., nodes  of in-degree zero)
and a single \textit{output node}  (i.e.,  a node of out-degree zero).  Input nodes are labeled with  either a variable or a field element in $\F$.
All the other nodes have in-degree two (unless otherwise
stated) and  are labeled by
either $+$ or $\times$. An input node is said to \emph{compute}  %
%
the variable or scalar that   labels  itself. A $+$ (or $\times$) gate
is said to compute the addition (product, resp.) of the polynomials
computed by its incoming nodes. An algebraic circuit is called a \emph{formula}, if the underlying directed acyclic graph  is a tree (that is, every node has at most one outgoing edge). The \emph{size} of a circuit is the number of nodes in it,
and the \emph{depth} of a circuit is the length of the longest directed path in it. 


\para{Algebraic Complexity Classes}
We  now recall some  basic notions from algebraic
complexity (for more details see \cite[Sec.~1.2]{SY10}).  
Over a ring $R$, $\cc{VP}_{R}$ (for
``Valiant's \P'') is the class of families $f=(f_n)_{n=1}^{\infty}$ of formal polynomials $f_n$ such that $f_n$ has $\poly(n)$ input variables, is of $\poly(n)$ degree, and can be computed by algebraic circuits over $R$ of $\poly(n)$ size. $\cc{VNP}_{R}$ (for ``Valiant's
\NP'') is the class of families $g$ of polynomials $g_n$ such that $g_n$ has $\poly(n)$ input variables and is of $\poly(n)$ degree, and can be written as
\[
g_n(x_1,\dotsc,x_{\poly(n)}) = \sum_{\vec{e} \in \{0,1\}^{\poly(n)}} f_n(\vec{e}, \vec{x})
\]
for some family $(f_n) \in \cc{VP}_{R}$.

A polynomial $f(\vec{x})$ is a \definedWord{projection} of a polynomial $g(\vec{y})$ if $f(\vec{x}) = g(L(\vec{x}))$ identically as polynomials in $\vec{x}$, for some map $L$ that assigns to each $y_i$ either a variable or a constant. A family of polynomials $(f_n)$ is a polynomial projection or \definedWord{p-projection} of another family $(g_n)$ if there is a function $t(n) = n^{\Theta(1)}$ such that $f_n$ is a projection of $g_{t(n)}$ for all (sufficiently large) $n$. The \emph{permanent} polynomial $\sum_{\sigma\in
S_n} \prod_{i=1}^nx_{i,\sigma(i)}$ (for $S_n$
the permutation group on $n$ elements) is complete under p-projections for $\VNP$. The \emph{determinant} polynomial on the other hand is known to be in
\VP\ but is not known to be complete for $\cc{VP}$ under p-projections. 


%
%
%
%

Two central questions in algebraic complexity theory are  whether the permanent is a p-projection of the determinant (a stronger variant speaks about quasi-polynomial projections); and  whether $\cc{VP}$ equals $\cc{VNP}$ \cite{Val79:ComplClass,Val79-permanent,Val82}. Since the permanent is complete for $\VNP$ (under p-projections), showing \VP$\neq$\VNP\ amounts to proving  that the permanent cannot be computed by  polynomial-size algebraic circuits. 

\subsection{Algebraic Proof Systems}
Let us now describe several algebraic proof systems for propositional logic (i.e.~for boolean tautologies). Assume we start from a set of initial polynomials (called \emph{axioms}) $f_1,\ldots,f_m\in\F[x_1,\ldots,x_n]$ over some field $\F$, then (the weak version of) Hilbert's Nullstellensatz shows that   $f_1(\vx)=\cdots=f_m(\vx)=0$ is unsatisfiable (over the algebraic closure of $\F$) if and only if there are polynomials $g_1,\ldots,g_m\in\F[\vx]$ such that $\sum_j g_j(\vx)f_j(\vx)=1$ (as a formal identity), or equivalently, that 1 is in the ideal generated by the $\{f_j\}_j$.

Beame, Impagliazzo, \Krajicek, Pitassi, and \Pudlak~\cite{BeameIKPP96} suggested to treat these $\{g_j\}_j$ as a \emph{proof} of the unsatisfiability of these axioms, called a \emph{\textbf{Nullstellensatz refutation}}.  This is particularly  relevant for complexity theory as one can restrict attention to \emph{boolean} solutions to these axioms by adding the \emph{boolean axioms}, that is, adding the polynomials $\{x_i^2-x_i\}_{i=1}^n$ to the axioms.  As such, one can then naturally encode $\NP$-complete problems such as the satisfiability of 3CNF formulas as the satisfiability of a collection  of constant-degree polynomials, and a Nullstellensatz refutation is then an equation of the form 
$$\sum_{j=1}^m g_j(\vx)f_j(\vx)+\sum_{i=1}^n h_i(\vx)(x_i^2-x_i)=1$$ for $g_j,h_i\in\F[\vx]$.  This proof system is sound and complete for refuting unsatisfiable axioms over $\bits^n$.
Given that the above proof system is sound and complete, it is then natural to ask what is its power to refute unsatisfiable collections of polynomial equations over $\bits^n$.  To understand this question one must define the notion of the \emph{size} of the above refutations.  Two popular notions are that of the \emph{degree}, and the \emph{sparsity} (number of monomials). 

Strong (linear) lower bounds on Nullstellensatz
degrees as well as strong (exponential) lower bounds on the sparsity of Nullstellensatz refutations are known (cf.~\cite{BeameIKPP96,BussIKPRS96,Razborov98,Grigoriev98,IPS99,BussGIP01,AlekhnovichRazborov01} and references therein). Unfortunately, the hard examples used for these lower bounds
\emph{do} admit polynomial-size proofs in stronger proof systems like Frege. 

Therefore, to correspond more accurately to Frege, strong algebraic proof systems must use a more economical representation of polynomials in proofs than  sum of monomials  (similarly to the way a boolean
formula is a much more succinct representation of a boolean
function than a mere CNF). The natural way is to measure the size of a polynomial by the size of the minimal algebraic circuit or formula
that computes it.

The idea to consider algebraic circuit size of algebraic proofs was raised initially by  Pitassi~\cite{Pit97} for Nullstellensatz written as
algebraic circuits, and was investigated further in   
\cite{GH03,RT06,RT07,Tza11-I&C} in the context of the polynomial calculus proof system.

Recently, Grochow and Pitassi~\cite{GrochowPitassi14} have suggested the following algebraic proof system that resembles the  Nullstellensatz, but with a variant that proved to have important consequences. A proof in the  Ideal Proof System is given as  a \uline{\textbf{\emph{single}}} polynomial, lending itself quite directly to algebraic circuit complexity techniques. In what follows we follow the notation in \cite{FSTW16}: 


\begin{definition}[Ideal Proof System (IPS), 
Grochow-Pitassi~\cite{GrochowPitassi14}]\label{def:orig-IPS} Let $f_1(\vx),\ldots,f_m(\vx)\in\F[x_1,\ldots,x_n]$ be a collection of polynomials. An \demph{IPS refutation} for showing that the polynomials $\{f_j\}_j$ have no common solution in $\bits^n$ is an algebraic circuit $C(\vx,\vy,\vz)\in\F[\vx,y_1,\ldots,y_m,z_1,\ldots,z_n]$, such that
        \begin{enumerate}
                \item $C(\vx,\vnz,\vnz) = 0$.
                \item $C(\vx,f_1(\vx),\ldots,f_m(\vx),x_1^2-x_1,\ldots,x_n^2-x_n)=1$.
        \end{enumerate}
        The \demph{size} of the IPS refutation is the size of the circuit $C$. If $C$ is of individual degree $\le 1$ in each $y_j$ and $z_i$, then this is a \demph{linear} IPS refutation (called basically \emph{Hilbert} IPS by Grochow-Pitassi~\cite{GrochowPitassi14}), which is  abbreviated as \lIPS. 
%
If $C$ comes from a restricted class of algebraic circuits $\cC$, then this is called a $\cC$-IPS refutation, and further called a $\cC$-\lIPS refutation if $C$ is linear in $\vy,\vz$. The variables $\vy,\vz$ are sometimes called the \emph{placeholder} \emph{variables} since they use as a placeholder for the axioms.    
\end{definition}

Notice that the definition above adds the equations $\{x_i^2-x_i\}_i$ to the system $\{f_j\}_j$.  It is \textit{not
}necessary (for the sake of completeness) to  add the equations $\vx^2-\vx$ to the system in general, but this is the most interesting regime for proof complexity and thus we adopt it as part of our definition. Also, note that the first equality in the definition of IPS means that the polynomial computed by $C$ is in the ideal generated by $\overline y,\overline z$, which in turn, following the second equality, means that $C$ witnesses the fact that $1$ is in the ideal generated by $f_1(\vx),\ldots,f_m(\vx),x_1^2-x_1,\ldots,x_n^2-x_n$ (the existence of this witness, for unsatisfiable set of polynomials, stems from the Nullstellensatz theorem as discussed above).


It is not hard to show that \lIPS\ is polynomially equivalent to the Nullstellensatz system, when both are measured by their \emph{circuit size}. For if we have an \lIPS\ refutation $C(\vx,\vy,\vz)$ we can turn it into a Nullstellensatz refutation by writing it as a sum of products of the (linear) variables $\vy,\vz$, with only a quadratic increase in size. For instance, if we write  $\vz'$ to denote $\vz$ without  $z_n$, we have $C(\vx,\vy,\vz)=C(\vx,\vy,\vz',z_n) = C(\vx,\vy,\vz',0)+ (C(\vx,\vy,\vz',1) - C(\vx,\vy,\vz',0))\cd z_n$. Now, since $C(\vx,\vy,\vz',0)$ does not contain the variable  $z_n$  we can continue in a similar way to ``take out'' the rest of the variables in $\vy,\vz$, one by one, reaching a Nullstellensatz refutation (when substituting the $f_i$'s and the boolean axioms for the $\vy,\vz$, respectively).      

Furthermore, Forbes, Shpilka, Tzameret and Wigderson \cite{FSTW16} showed that \lIPS\ refutations written as  (general) algebraic circuits is \textit{polynomially equivalent} to IPS (though for restricted classes $\cC$, $\cC$-IPS may differ
 from $\cC$-\lIPS).

Considering  both the Nullstellensatz and the IPS we can see that the main innovation in the IPS is the introduction of the placeholder variables $\overline y,\overline z$. This idea enables considering a refutation as a \emph{single} polynomial instead of considering a collection of polynomials (that is, those polynomial coefficients of the initial axioms, as is  the case of the Nullstellensatz).  
\smallskip

Grochow-Pitassi~\cite{GrochowPitassi14} showed that the IPS system is very powerful and can simulate Extended Frege (this follows from the fact that IPS is a generalization of the Nullstellensatz written as algebraic circuits and already \cite{Pit97} showed that the latter system simulates Extended Frege).

The fact that $\cC$-IPS refutations are efficiently checkable (with randomness) follows from the fact that we only need to verify the polynomial identities stipulated by the definition.  That is, it suffices to solve an instance of the \emph{\textbf{polynomial identity testing} (PIT)} problem for the class $\cC$: given a circuit from the class $\cC$ decide whether it computes the identically zero polynomial.  This problem is solvable in probabilistic polynomial time ($\BPP$) for general algebraic circuits, and there are various restricted classes for which deterministic algorithms are known (see \autoref{sec:PIT}). \medskip

\para{The Polynomial Calculus}

The Polynomial Calculus is an  algebraic proof system introduced by \cite{CleggEI96}. It can be considered as a ``dynamic'' version of the Nullstellensatz; namely, instead of providing a single certificate that 1 is in the ideal of the initial (unsatisfiable) polynomials, in PC we are allowed to derive the polynomial 1 step by step, by working in the ideal generated by the initial polynomials.

\begin{definition}[\demph{Polynomial Calculus (PC)}]\label{def:PC}
        Let $\mathbb{F}$ be a field and let $F= \{f_1,\ldots,f_m\}$ be a collection of multivariate polynomials from $\mathbb{F}[x_1,\ldots, x_n]$.
A \emph{PC proof from  $Q$ of a polynomial $g$} is a finite sequence $\pi =(p_1 ,\ldots,p_\ell)$ of multivariate polynomials from $\mathbb{F}[x_1,\ldots, x_n]$, where $p_\ell=g$ and for every $ 1\le i\le \ell$, either $p_i = f_j\,$ for some $j\in[m]$, or $p_i$ is a boolean axiom $x_i\cdot(1-x_i)$ for some $i\in[n]$,  or $p_i$ was derived from $p_j,p_k\,$, for  $j,k<i$, by one of the following inference rules:
\begin{enumerate}
\item[(i)] \emph{Product rule}: from $p$, derive $x_i\cd p$, for $i\in[n]$;

\item[(ii)] \emph{Addition rule}: from $p,q$, derive $a p + b q$, for $a,b \in \F$.
\end{enumerate}
A \emph{{PC refutation} of} $F$ is a proof of $\;1$ (which is interpreted as $1=0$, that is the unsatisfiable equation standing for {\rm\textsf{false}}) from $F$.
\end{definition}

Similar to the Nullstellensatz, the standard complexity measures for PC are the \emph{degree} of a PC proof, which is the maximal (total) degree of a polynomial in the proof and the \emph{size} of a PC proof which is the total number of monomials (with nonzero coefficients) in all the PC proof lines. However, it is also possible to consider the total algebraic circuit size of  all the  PC proofs lines as a complexity measure.
\smallskip 

\para{Non-commutative Algebraic Proof Systems}

Motivated by the fact that the class of non-commutative formulas admits a deterministic PIT algorithm by Raz and Shpilka \cite{RazShpilka05}, and even more importantly admits  exponential-size lower bounds by Nisan \cite{Nisan91}, Li, Tzameret and Wang~\cite{LTW15-CCC} considered a variant of the  IPS over \emph{non-commutative} polynomials written as non-commutative formulas. Their non-commutative
IPS was shown to constitute a tighter characterization of Frege proofs than the original (commutative) IPS: first, proofs in this system are checkable in \textit{deterministic} polynomial-time; and second, Frege can simulate (with a quasi-polynomial increase in size) non-commutative IPS refutations (over the field of  two elements). But perhaps most importantly, the fact that we do have lower bounds on non-commutative formulas together with the  characterization of  any Frege proof as a single non-commutative formula, gives some hope to progress on the problem of Frege lower bounds. We discuss the non-commutative IPS in more details in  Section \ref{sec:noncommIPS}.

\section{IPS}\label{sec:IPS}

In this section we show that lower bounds for IPS
imply algebraic circuit lower bounds, namely that the permanent
does not have polynomial-size algebraic circuits. This implication
is interesting because it is a unique case in proof complexity where a lower bound on a \emph{specific} proof system (on any tautology) is shown to imply explicit circuit lower bounds. We then compare the strength of IPS to Extended Frege. We show that IPS, in its full generality polynomially simulates Extended Frege, and on the other hand, show that Extended Frege polynomially simulates IPS if PIT  has feasible correctness proofs in Extended Frege.\footnote{Namely, that Extended Frege has polynomial-size proofs of the statement expressing that the PIT for algebraic circuits is decidable by polynomial-size Boolean circuits.}

\subsection{Lower Bounds on IPS Imply Algebraic Circuit Lower Bounds} \label{sec:VNP}
\begin{theorem}[Grochow-Pitassi~\cite{GrochowPitassi14}] \label{thm:VNP} 
For any ring $R$, a super-polynomial lower bound on 
IPS proofs over $R$ of any family of tautologies implies 
$\mathsf{VNP}_{R} \neq \mathsf{VP}_{R}$.
A super-polynomial lower bound on the number of proof-lines in polynomial calculus proofs implies that the permanent is not a p-projection of the determinant. 
\end{theorem}

We will sketch the proof for the first half of the  theorem which
gives the main idea. The proof of the second half can be found in
\cite{GrochowPitassi14}.

\begin{lemma} \label{lem:VNP}
Every family of unsatisfiable  CNF formulas  $(\varphi_n)$ has a family of IPS certificates $(C_n)$ in $\mathsf{VNP}_{R}$.
\end{lemma}


\noindent\textit{Proof of Theorem~\ref{thm:VNP}, assuming Lemma~\ref{lem:VNP}.}
Our proof  is taken from \cite{GrochowPitassi14}.
For a given set $\mathcal{F}$ of unsatisfiable polynomial equations $F_1=\dotsb=F_m=0$, a lower bound on IPS 
refutations of $\mathcal{F}$ is equivalent to giving the same circuit lower bound on \emph{all} IPS certificates for 
$\mathcal{F}$. A super-polynomial lower bound on 
IPS implies that some function in $\mathsf{VNP}$---namely, 
the $\mathsf{VNP}$-IPS certificate guaranteed by Lemma~\ref{lem:VNP}---cannot be computed by\ polynomial-size algebraic circuits, and hence that $\mathsf{VNP} \neq \mathsf{VP}$. 
\qed


\medskip 

\noindent\textit{Proof sketch of Lemma~\ref{lem:VNP}.}
We mimic one of the proofs of completeness for linear
 IPS \cite[Theorem~1]{Pit97} 
and then show that this
proof can in fact be carried out in $\mathsf{VNP}$. We omit any mention of the ground ring, as it will not be relevant.

Let $\varphi_n(\vec{x}) = \kappa_1(\vec{x}) \wedge \dotsb \wedge \kappa_m(\vec{x})$ be an unsatisfiable CNF formula, where each $\kappa_i$ is a disjunction of literals. Let $C_i(\vec{x})$ denote the (negated) polynomial translation of $\kappa_i$ via $\neg x \mapsto x$, $x \mapsto 1-x$ and $f \vee g \mapsto fg$; in particular, $C_i(\vec{x}) = 0$ if and only if $\kappa_i(\vec{x}) = 1$, and thus $\varphi_n$ is unsatisfiable if and only if the system of equations $C_1(\vec{x})=\dotsb=C_m(\vec{x})=x_1^2 - x_1 = \dotsb = x_n^2 - x_n = 0$ is unsatisfiable. In fact, as we will see in the course of the proof, we will not  need the equations $x_i^2 - x_i = 0$. It will be convenient to introduce the function $b(e,x)=ex + (1-e)(1-x)$, i.e., $b(1,x) = x$ and $b(0,x)=1-x$. For example, the clause $\kappa_i(\vec{x}) = (x_1 \vee \neg x_{17} \vee x_{42})$ gets translated into $C_i(\vec{x}) = (1-x_1)x_{17}(1-x_{42}) = b(0,x_1)b(1,x_{17})b(0,x_{42})$, and therefore an assignment falsifies $\kappa_i$ if and only if $(x_1,x_{17},x_{42}) \mapsto (0,1,0)$.

Just as $1 = x_1 x_2 + x_1(1-x_2) + (1-x_2)x_1 + (1-x_2)(1-x_1)$, an easy induction shows that
\begin{equation} \label{eqn:1}
1 = \sum_{\vec{e} \in \{0,1\}^{n}} \prod_{i=1}^{n}b(e_i,x_i).
\end{equation}
We will show how to turn this expression into a $\mathsf{VNP}$ certificate refuting $\varphi_n$. Let $c_i$ be the placeholder variable corresponding to $C_i(\vec{x})$.

The idea is to partition the assignments $\{0,1\}^{n}$ into $m$ parts $A_1,\dotsc,A_m$, where all assignments in the $i$-th part $A_i$ falsify clause $i$. This will then allow us to rewrite equation (\ref{eqn:1}) as
\begin{equation} \label{eqn:rewrite}
1 = \sum_{i=1}^{m} C_i(\vec{x})\left(\sum_{\vec{e} \in A_i} \prod_{j : x_j \notin \kappa_i} b(e_j,x_j)\right),
\end{equation}
where ``$x_j \notin \kappa_i$'' means that neither $x_j$ nor its negation appears in $\kappa_i$. Equation (\ref{eqn:rewrite}) then becomes the IPS-certificate $\sum_{i=1}^{m} c_i \cdot (\sum_{\vec{e} \in A_i} \prod_{j : x_j \notin \kappa_i} b(e_j,x_j))$. What remains is to show that the sum can indeed be rewritten this way, and that there is some partition $(A_1,\dotsc, A_m)$ as above such that the resulting certificate is in fact in $\mathsf{VNP}$.

First, let us see why such a partition allows us to rewrite (\ref{eqn:1}) as (\ref{eqn:rewrite}). The key fact here is that the clause polynomial $C_i(\vec{x})$ divides the term $t_{\vec{e}}(\vec{x}) := \prod_{i=1}^{n} b(e_i, x_i)$ if and only if $C_i(\vec{e}) = 1$, if and only if $\vec{e}$ \emph{falsifies} $\kappa_i$. Let $C_i(\vec{x})=\prod_{i \in I} b(f_i,x_i)$, where $I \subseteq [n]$ is the set of indices of the variables appearing in clause $i$. By the properties of $b$ discussed above, $1=C_i(\vec{e})=\prod_{i \in I} b(f_i, e_i)$ if and only if $b(f_i,e_i)=1$ for all $i \in I$, if and only if $f_i=e_i$ for all $i \in I$. In other words, if $1=C_i(\vec{e})$ then $C_i = \prod_{i \in I} b(e_i, x_i)$, which clearly divides $t_{\vec{e}}$. Conversely, suppose $C_i(\vec{x})$ divides $t_{\vec{e}}(\vec{x})$. Since $t_{\vec{e}}(\vec{e})=1$ and every factor of $t_{\vec{e}}$ only takes on boolean values on boolean inputs, it follows that every factor of $t_{\vec{e}}$ evaluates to $1$ at $\vec{e}$, in particular $C_i(\vec{e})=1$.

Let $A_1, \dotsc, A_m$ be a partition of $\{0,1\}^n$ such that every assignment in $A_i$ falsifies $\kappa_i$. Since $C_i$ divides every term $t_{\vec{e}}$ such that $\vec{e}$ falsifies clause $i$, $C_i$ divides every term $t_{\vec{e}}$ with $\vec{e} \in A_i$, and thus we can indeed rewrite (\ref{eqn:1}) as (\ref{eqn:rewrite}).

Next, we show how to construct a partition $A_1, \dotsc, A_m$ as above so that the resulting certificate is in $\mathsf{VNP}$. The partition we will use is a greedy one. $A_1$ will consist of \emph{all} assignments that falsify $\kappa_1$. $A_2$ will consist of all \emph{remaining} assignments that falsify $\kappa_2$. And so on. In particular, $A_i$ consists of all assignments that falsify $\kappa_i$ and \emph{satisfy} all $A_j$ with $j < i$. (If at some clause $\kappa_i$ before we reach the end, we have used up all the assignments---which happens if and only if the first $i$ clauses on their own are unsatisfiable---that's okay: nothing we've done so far nor anything we do below assumes that all $A_i$ are nonempty.)

Equivalently, $A_i = \{\vec{e} \in \{0,1\}^n \;|\; C_i(\vec{e})=1 \text{ and } C_j(\vec{e})=0 \text{ for all } j < i\}$. For any property $\Pi$, we write $\llbracket \Pi(\vec{e}) \rrbracket$ for the indicator function of $\Pi$: $\llbracket \Pi(\vec{e}) \rrbracket=1$ if and only if $\Pi(\vec{e})$ holds, and $0$ otherwise. We thus get the certificate:
\begin{eqnarray*}
& & \sum_{i=1}^{m} c_i \cdot \left(\sum_{\vec{e} \in \{0,1\}^n} \llbracket \vec{e} \text{ falsifies } \kappa_i \text{ and satisfies $\kappa_j$ for all } j < i \rrbracket \prod_{j : x_j \notin \kappa_i} b(e_j, x_j) \right) \\
& = & \sum_{i=1}^{m} c_i \cdot \left(\sum_{\vec{e} \in \{0,1\}^n} \llbracket C_i(\vec{e})=1 \text{ and } C_j(\vec{e})=0 \text{ for all } j < i \rrbracket \prod_{j : x_j \notin \kappa_i} b(e_j, x_j) \right) \\
& = & \sum_{i=1}^{m} c_i \cdot \left(\sum_{\vec{e} \in \{0,1\}^n} \left(C_i(\vec{e}) \prod_{j < i}(1-C_j(\vec{e})) \right) \prod_{j : x_j \notin \kappa_i} b(e_j, x_j) \right) \\
& = & \sum_{e \in \{0,1\}^{n}} \sum_{i=1}^{m} c_i C_i(\vec{e})\left(\prod_{j < i}(1-C_j(\vec{e}))\right)\left(\prod_{j : x_j \notin \kappa_i} b(e_j, x_j)\right)
\end{eqnarray*}
              
Finally, it is readily visible that the polynomial function of $\vec{c}$, $\vec{e}$, and $\vec{x}$ that is the summand of the outermost sum $\sum_{\vec{e} \in \{0,1\}^{n}}$ is computed by a polynomial-size circuit of polynomial degree, and thus the entire certificate is in $\mathsf{VNP}$. 
\qed
                                                                                                                                  
\subsection{IPS  Polynomially Simulates Extended Frege}

In this section we show that IPS  polynomially simulates Extended Frege. For simplicity, we exemplify this simulation by showing how
IPS written as algebraic \emph{formulas} polynomially simulates the
Frege proof system, but the proof for Extended Frege is quite similar.
It was further shown by \cite{GrochowPitassi14} that restricted subsystems of IPS can polynomially simulate the corresponding restricted subsystem of Extended Frege, and specifically this holds for IPS written as constant-depth algebraic circuits and constant-depth Frege systems with modulo counting gates ($\AC^0[p]$-Frege).

\begin{theorem}[Grochow-Pitassi~\cite{GrochowPitassi14}]\label{thm:IPS-simulates-EF}
Let $\varphi$ be a 3CNF formula.  If there is an Extended Frege proof (Frege proof) that $\varphi$ is unsatisfiable in size-$s$, then there is an IPS refutation of circuit (formula, resp.) size $\poly(|\varphi|,s).$ 
\end{theorem}
\noindent\textit{Proof sketch.} One way of thinking of this simulation (and similar simulations of propositional systems by IPS-variants) is to consider a two-step conversion of propositional proofs into IPS refutations as follows. First, turn the Frege refutation  into a tree-like Frege refutation and convert each proof-line in the tree-like Frege refutation into an equivalent algebraic formula, obtaining a tree-like PC refutation. Secondly, convert the tree-like PC proof into a single formula, whose underlying formula-tree is precisely the underlying tree of the PC proof.

Note that a Frege proof can be converted into a \textit{tree-like proof} with only a polynomial increase in size, that is, a proof in which every proof-line can be used at most once in modus ponens (cf.~\cite{Kra95}). Therefore, we start from a tree-like Frege refutation of the unsatisfiable 3CNF formula (namely, a proof of \textsf{false} from the clauses of the 3CNF formula as assumptions), and show how to obtain from this  an IPS refutation of the arithmetic version (see below) of the same 3CNF formula. Thus, a Frege proof of \textsf{false} from a given CNF $\varphi$ is translated into an IPS proof of $1$ from the initial (arithmetic version of) $\varphi$, yielding an IPS refutation of $\varphi$. 
\smallskip 

\emph{\uline{Step 1}}: This step involves the arithmetization of Frege proofs, namely, converting each Frege proof-line to an algebraic formula. The transformation converts a boolean formula into a corresponding algebraic formula (over the rationals, or $\F_q$, for a prime $q$; the simulation uses only the fact that the field has $1,0$ and $-1$) that evaluates to  $0$ for all $0$-$1$ assignments. This is done in  the same way as in the proof of Lemma \ref{lem:VNP}: \textsf{true} becomes 0, \textsf{false} becomes 1, a variable $x_i$ becomes $1-x_i$, $\neg A$ becomes $1-\tr(A)$, where $\tr(A)$ denotes the translation of $A$, $A\lor B$ becomes the product of the corresponding translations $\tr(A)\cd \tr(B)$, and $A\land B$ becomes $1-(1-\tr(A))\cd(1-\tr(B))$ (Schoenfield's system uses the $\to$ logical connective,
but we can simply treat $A\to B$ as an abbreviation of $\neg A\lor B$). 
It is easy to check that for any 0-1 assignment, $A$ evaluates to \textsf{true} iff $\tr(A)=0$.

Once we converted every Frege proof-line into its corresponding algebraic formula, we get something that \textit{resembles} a sequential algebraic proof, namely a PC proof. However, it is not precisely a  legitimate PC proof because, e.g., every application of modus ponens (from $A$ and $A\to B$ derive $B$) is translated into the purported rule \emph{``from $\tr(A)$ and $(1-\tr(A))\cd \tr(B)$ derive  $\tr(B)$}'', which is not a formal rule in PC. Nevertheless, we can make this arithmetized Frege proof into a legitimate PC proof, except that our PC proof will have a \emph{generalized} product rule: instead of being able to multiply a proof-line only by a \emph{single }variable we will enable a product by a \emph{polynomial}, namely, from $f$ derive $g\cd f$, for some polynomial $g\in\F[x_1,\ldots,x_n]$. 

To form our (generalized) PC proof  we simply \emph{simulate} Schoenfield's system rules and axioms. Considering the  example above, we need to construct a short PC proof of $\tr(B)$ from $\tr(A)$ and $(1-\tr(A))\cd \tr(B)$: first derive  $\tr(A)\cd \tr(B)$ by the generalized PC product rule and then add this  to $(1-\tr(A))\cd \tr(B),$ to obtain $\tr(B)$. Similarly, Frege axioms are translated into algebraic formulas, that we then need to \emph{derive} in PC, and this is  possible to do efficiently. \smallskip

\uline{\emph{Step 2}}: Here we transform the (generalized) PC refutation  from step 1, whose underlying proof-graph is a tree (since we assumed without loss of generality that our initial Frege proof is a tree-like proof), into a \emph{single formula} whose underlying graph is essentially the same tree. This formula constitutes the IPS refutation  of  the arithmetic translation  of $\varphi$. The transformation from a PC proof to a formula is quite straightforward. For example, assume that in the PC proof we derived $g\cd f$ from $f$. And suppose that we already  built the IPS proof of $f$, namely $C(\vx,f_1(\vx),\ldots,f_m(\vx),x_1^2-x_1,\ldots,x_n^2-x_n)=f$. Then, $g\cd C(\vx,f_1(\vx),\ldots,f_m(\vx),x_1^2-x_1,\ldots,x_n^2-x_n)=g\cd f$ is the IPS proof of $g\cd f$. Simulating the addition rule of PC is done in a similar manner. (Formally, the $f_i(\overline x)$'s should be substituted by the placeholder variables $\overline y$, and the boolean axioms by the placeholder variables $\overline z$.)  

It is easy to see that the resulted IPS is of size polynomial in the size of the PC refutation, which in turn is of size polynomial in the size of the original Frege refutation .
\qed



\subsection{PIT as a Bridge Between Circuit Complexity and Proof Complexity} \label{sec:PIT_Bridge}

In this section we 
sketch the argument that Extended Frege (EF) 
is polynomially equivalent to IPS if there are polynomial-size circuits for PIT whose correctness---suitably formulated---can be efficiently proved in EF.
More precisely, we identify a small set of natural axioms for PIT and show that if these axioms can be proven efficiently in EF, then EF is p-equivalent to IPS. 

The high-level idea is to formalize soundness of IPS as a
sequence of propositional statements and then to show: 
\begin{itemize}
\item[(1)] if
EF has efficient proofs of IPS soundness then EF can polynomially
simulate IPS; 
\item[(2)] Show that EF has efficient proofs of IPS soundness
if a small set of natural axioms for PIT are efficiently provable in EF.
\end{itemize}

The idea behind (1) is not new and traces back to Hilbert;
its counterpart for propositional proof systems was first
formalized by Cook \cite{Coo75}. We explain the idea for propositional refutation
systems here.
Soundness of a propositional proof system 
states that any formula that has a proof (in the system)
is a tautology. Formalizing soundness propositionally involves
studying {\it partial} soundness, where we have a different propositional
formula for each proof length.
In more detail, for a propositional proof system $Q$, 
$Soundness_{Q,n}, n > 0$ will be a family of propositional statements. The underlying
variables of $Soundness_{Q,n}$ are $\vec{x}$, $\vec{y}$ and $\vec{z}$,
where we think of $\vec{x}$ as an encoding of some $Q$-proof
of length $n$, $\vec{y}$ as an encoding of
a $k$-DNF formula with $n' \leq n$ underlying variables, and
$\vec{z}$ as a boolean assignment to the $n'$ underlying variables.
$Soundness_{Q,n}(\vec{x},\vec{y},\vec{z})$ is
of the form
$Proof_{Q,n}(\vec{x},\vec{y}) \rightarrow Truth(\vec{y},\vec{z})$
where
$Proof_{Q,n}(\vec{x},\vec{y})$ expresses that $\vec{x}$ is an encoding
of a $Q$-proof of the formula encoded by  $\vec{y}$, 
and $Truth(\vec{y},\vec{z})$ expresses that $\vec{z}$
satisfies the formula encoded by $\vec{y}$ (i.e., the formula encoded by $\vec{y}$ is a tautology).

For sufficiently strong propositional
proof systems $P$ and $Q$, it is well-known that $P$ polynomially simulates $Q$
if and only if there are polynomial-sized $P$-proofs of $Soundness_{Q,n}$ for all $n > 0$.
The intuitive argument is as follows:
Suppose that $Q$ has a short proof of some formula $g$; let $\alpha(g)$ be the encoding
of $g$, and let $\beta(g)$ be the encoding of the short $Q$-proof of $g$.
Then since $P$ has short proofs of $Soundness_{Q,n}$, we instantiate this
with $g$ to give a short $P$-proof of $Soundness_{Q,n}(\beta(g),\alpha(g),\vec{z})$.
Since $Proof_{Q,n}(\beta(g),\alpha(g))$ is a tautology and involves no
propositional variables, it has a short $P$-proof and thus by modus ponens,
there is a short $P$-proof of $Truth(\alpha(g),\vec{z})$.
The last step is to demonstrate short $P$-proofs of $Truth(\alpha(g),\vec{z}) \rightarrow g$.

We will take $P$ to be EF and $Q$ to be IPS.
Then EF can polynomially simulate $Q$ if and only if EF can
efficiently prove the soundness tautologies for IPS.
Proving the soundness tautologies for IPS amounts to
stating and proving (in Extended Frege) that if $C$ is an algebraic
circuit such that:
(1) $C(\vec{x},\vec{0},\vec{0})=0$ and
(2) $C(\vec{x},f_1(\vec{x}),\ldots,f_m(\vec{x}))=1$, then
$f_1,\ldots,f_m$ is unsatisfiable.
In order to state (1) and (2) efficiently, we need polynomial-sized circuits for
polynomial identity testing. Then in order to prove that
(1) and (2) imply that $f_1,\ldots,f_m$ is unsatisfiable, we will
need to use basic properties of our PIT circuits.
We omit the proof here, but will informally state the axioms that will be required in order to carry out the above plan.
\vspace{-3pt} 
\subsection{Axioms for Circuits for Polynomial Identity Testing}\label{sec:PIT_Axioms}
Fix some standard boolean encoding of algebraic circuits, \mar{I assume here you are working over a polynomial sized field?
}
so that the encoding of any size-$m$ algebraic circuit has size 
$\poly(m)$. 
We use ``$[C]$'' to denote the encoding of the 
algebraic circuit $C$. 
Let $K = (K_{m,n})$ denote a family of boolean circuits for solving polynomial identity testing. That is, $K_{m,n}$ is a boolean function that takes as input the encoding of a size $m$ algebraic circuit, $C$, 
over variables $x_1, \ldots, x_n$, and if $C$ has polynomial degree, then $K$ outputs 1 if and only if the polynomial computed by $C$ is the 0 polynomial.



%
The first axiom states that if $C$ is a circuit over variables $\vec{x}$ computing
the identically 0 polynomial, then the circuit $C$ where we plug in a particular
boolean input $\vec{p}$, still computes the identically 0 polynomial:
$$K([C(\vec{x})]) \rightarrow K([C(\vec{p})]).$$
The second axiom states that if $C$ is a circuit over variables $\vec{x}$ computing
the zero polynomial, then the circuit $1-C$ does not compute the zero polynomial:
$$K([C(\vec{x})]) \rightarrow \neg K([1-C(\vec{x})]).$$
The third axiom states that 
if the polynomial computed by circuit
$G$ is 0, then $G$ can be substituted for the constant $0$:
$$K([G({\vec x})]) \land K([C({\vec x},0)]) \rightarrow K([C({\vec x},G({\vec x}))]).$$
Finally, the last axiom states that PIT is closed under
permutations of the variables. More specifically if $C(\vec{x})$ is identically 0,
then so is $C(\pi(\vec{x}))$ for all permutations $\pi$:



Note that the issue is not the existence of small circuits for PIT since we would be happy with nonuniform polynomial-size PIT circuits, which do exist. Unfortunately the known constructions are highly nonuniform---they involve picking random points---and we do not see how to prove the above axioms for these constructions. 
On the other hand, it is widely conjectured that there exist uniform polynomial-sized circuits
for PIT, and it is therefore a very intriguing question whether or not the proofs of correctness
of such uniform algorithms (assuming that they exist) can be carried out in a feasible (polynomial-time) proof system.

\section{Non-Commutative IPS }\label{sec:noncommIPS}
In this section we discuss the non-commutative IPS, introduced by Li, Tzameret and Wang \cite{LTW15-CCC}, which is a variant of the IPS over non-commutative polynomials. The main result is that the non-commutative IPS  completely captures (up to quasi-polynomial factors)
the Frege proof system when the non-commutative IPS refutations are written as non-commutative formulas.

Since the class of non-commutative formulas are well understood,
namely, it admits exponential-size lower bounds by Nisan \cite{Nisan91},
and deterministic PIT algorithm by Raz-Shpilka \cite{RazShpilka05},
this characterization of a Frege proof by a single non-commutative
formula gives some hope for better understanding of specific
Frege proofs and specifically for the eventual possibility of providing lower bounds on Frege proofs.
\smallskip 
 
We need to describe first the basic setup before giving the precise definition. A \textbf{\textit{non-commutative polynomial }}is a polynomial in which products are non-commuting, namely, $x_ix_j$ is not the same polynomial as $x_jx_i$, whenever $i\neq j$. In other
words, $x_i x_j-x_jx_i$ is not the zero polynomial. Thus, we can treat a non-commutative polynomial as a formal sum of non-commutative monomials. We denote by $\F\langle x_1,\ldots,x_n\rangle$ the ring of non-commutative polynomials over the variables $x_1,\ldots,x_n$. A \textit{\textbf{non-commutative formula}} is the same as a (commutative) algebraic formula only that the children of product gates have \textit{order}, so that we can record the order of multiplication. Therefore, the polynomial that a non-commutative formula computes is the polynomial achieved by first multiplying out  brackets whereby we get a sum of monomials in which the order of multiplication matters (without performing still any cancelations of monomials), and then performing monomial cancelation (and grouping) only when two monomials have the same variables with the same powers and \emph{the same order of multiplication}. 

It helps to think of non-commutative
polynomials (and formulas) as a means to compute functions over non-commutative
domains such as matrix algebras (in which matrix product is non-commuting in
general).


\begin{definition}[Non-commutative IPS, Li-Tzameret-Wang~\cite{LTW15-CCC}]
\label{def:non-commutative-IPS}
Let $\F$ be a field. Let $f_1(\vx),\ldots,f_m(\vx)\in\F\langle \vx\rangle$ be a system of non-commutative polynomials. A \demph{non-commutative-IPS refutation} that the polynomials $\{f_j\}_j$ have no common solution in $\bits^n$,\footnote{One can check that the $f_i(\overline x)$'s have no common $0$-$1$ solutions in $\F$ iff they do not have a common 0-1 solution in every $\F$-algebra.} is a  non-commutative formula  
$\mathfrak{F}(\vx,\vy,\vz,\vw)\in\F\langle\vx,y_1,\ldots,y_m,z_1,\ldots,z_n,w_1,\ldots,w_{n\choose 2}\rangle$, such that
\begin{enumerate}
                \item $\mathfrak{F}(\vx,\vnz,\vnz,\vnz) = 0$.
                \item $\mathfrak{F}(\vx,f_1(\vx),\ldots,f_m(\vx),\vx^2-\vx,x_1x_2-x_2x_1,\ldots,x_{n-1}x_n-x_nx_{n-1})=1$.
        \end{enumerate}
The $\vx^2-\vx$ denotes
the boolean axioms $x_i^2-x_i$, for all $i\in[n]$, and $x_ix_j-x_jx_i$, for all $i<j\in[n]$,
 are called the \emph{commutator}
\emph{axioms}. The \demph{size} of a non-commutative IPS refutation is the minimal size of a \emph{non-commutative formula} computing the non-commutative-IPS refutation. 

\end{definition}

The novelty in the non-commutative IPS  in comparison
to the original (commutative) IPS is simply that a
single refutation is   a non-commutative
polynomial instead of a commutative one.

One way of thinking about a non-commutative IPS refutation is as a \emph{commutative} IPS formula augmented with \emph{additional proofs} for  demonstrating that all the monomials computed along the
way in this formula are indeed commuting. More precisely, consider a \emph{commutative} IPS refutation  written as a formula $F(\vx,\vy,\vz)$, such that $F(\vx,f_1(\vx),\ldots,f_m(\vx),\vx^2-\vx)=1$
as a \emph{commutative} formula but $F(\vx,f_1(\vx),\ldots,f_m(\vx),\vx^2-\vx)\neq
1$ as a non-commutative formula. In the commutative version of   $F(\vx,f_1(\vx),\ldots,f_m(\vx),\vx^2-\vx)$,
two monomials computed by this refutation,
say $x_1x_2x_3$ and $x_2x_1x_3$, will be considered the same
monomial. However, in the non-commutative version these two monomials are
considered distinct, so we need to add an explicit
proof that $x_1x_2x_3$ is equal to $x_2x_1x_3$---in this case the proof added is
simply  $(x_1x_2-x_2x_1)\cd x_3$ which is a right product of a commutator axiom. 

 Note that to
achieve the \emph{completeness} of the system we \emph{must} add the commutator axioms.
Indeed, the non-commutative polynomial $1+x_ix_j-x_jx_i$,
for example, is unsatisfiable over 0-1 solutions, but it cannot be proven unsatisfiable without using the commutator axioms, because it \emph{is} satisfiable over some non-commutative matrix algebra (and by soundness of the non-commutative IPS there cannot be a proof of its unsatisfiability). 

The gist of Li-Tzameret-Wang's simulation of Frege by the non-commutative IPS is that even when we add the commutator
axioms, and by that  force each refutation to explicitly witness any cancelation between (commuting) monomials, we are still not weakening the system too much, namely, we still keep the system as strong as the Frege system. The reason for this is that in Frege we consider propositional formulas as purely \textit{syntactic} terms. For example, if  $F[z]$ is a propositional formula,  then $F[(A\land B)/z]$ and $F[(B\land A)/z]$ (which are the results of substituting $A\land B$ and $B\land A$ for $z$ in $F$, resp.) are two \emph{different} formulas and the tautology  $F[(A\land B)/z]\equiv F[(B\land A)/z]$ requires an explicit Frege proof.

Non-commutative IPS polynomially
simulates Frege, and conversely, Frege quasi-polynomially simulates non-commutative
IPS over $GF(2)$ (for the latter see next section):

\begin{theorem}[Li-Tzameret-Wang~\cite{LTW15-CCC}]\label{thm:LTW}
Let $\varphi$ be an unsatisfiable propositional formula.  If Frege can prove that $\varphi$ is unsatisfiable in size-$s$, then there is a non-commutative IPS refutation of formula size $\poly(|\varphi|,s)$ computing a polynomial of degree $\poly(|\varphi|,s)$.  Further, this refutation is checkable in deterministic $\poly(|\varphi|,s)$ time. 
\end{theorem}

The idea to consider non-commutative
formulas in algebraic proofs as well as adding the
commutator axioms was considered originally
  by Tzameret 
\cite{Tza11-I&C}, though in that work the proof system
was built on the polynomial calculus and not the IPS, and therefore did not obtain the characterization of a Frege proof as a single 
non-commutative formula.

Let us sketch the proof of \ref{thm:LTW}. We begin with the simulation of Frege by non-commutative IPS. The idea here is quite similar to the simulation of Frege by
(formula) IPS (Theorem \ref{thm:IPS-simulates-EF}).

\medskip 
\noindent\textit{Non-commutative IPS polynomially simulates
Frege (proof sketch)}.
Let us consider, as in the proof of Theorem  \ref{thm:IPS-simulates-EF},
a two-step simulation of Frege by non-commutative IPS.
We start from a Frege proof of \textsf{false} from the formula $\phi$ serving as an assumption in the proof, that we assume without loss of generality is a tree-like proof.  

\uline{\textit{Step 1}}: Here
we convert
each proof-line into an algebraic formula  in the same way we did
in the proof of Theorem \ref{thm:IPS-simulates-EF}, using the same translation function $\tr(\cd)$, only  now let  $tr(\cd)$
 return a  \emph{non-commutative
formula}. So, for instance,
assuming $A$ and $B$ are unequal,
 $\tr(A\lor B) = \tr(A)\cd \tr(B)\neq
\tr(B)\cd\tr(A)=\tr(B\lor A)$ 
(note that any algebraic formula can represent
either a commutative or a non-commutative polynomial; namely, a non-commutative
formula computes a non-commutative polynomial by taking into account
the order in which children of product gates  appear in the formula).

Now, as before, we get a purported refutation  of $\tr(\varphi)$ that only resembles a PC refutation, 
and in addition the polynomials
are non-commutative. We wish to complement this purported  proof into a legitimate
algebraic proof operating
with non-commutative polynomials;
this will be in fact the \emph{non-commutative PC} system defined by Tzameret
\cite{Tza11-I&C}: it is similar to the  PC proof system, only that polynomials are considered as non-commutative polynomials,
the addition rule is the same as in PC, and the generalized product rule can be applied either from the right or from the left, namely, from $f$ derive either $g\cd f$ or $f\cd g$, for some $g$; further, in addition to the boolean axioms, we add the commutator axioms $x_ix_i-x_jx_i$,
for every pair of variables, to the system.

We consider the case of simulating the first axiom of Schoenfield's system $A\to(B\to A)$ in this non-commutative PC system. This will exemplify why we need to use the commutator axioms. Thus, consider the translation of this axiom under $\tr(\cd)$. Recall that   $\to$ is just an abbreviation. Then, $\tr(A\to(B\to A))=\tr(\neg A \lor (\neg B\lor A))=(1-\tr(A))\cd((1-\tr(B))\cd \tr(A))$. Our goal is to construct a non-commutative PC proof of the following \emph{non-commutative} polynomial: 
\begin{equation}\label{eq:non-commutative_Modus_ponens}
(1-\tr(A))\cd((1-\tr(B))\cd
\tr(A))\,.
\end{equation} 
For this purpose, we first derive the polynomial
$\tr(A)-\tr(A)^2=(1-\tr(A))\cd \tr(A)$. This is doable efficiently using only the boolean axioms $x_i-x_i^2$ (by induction on the size of $A$). Then, we wish to derive \eqref{eq:non-commutative_Modus_ponens}
from $(1-\tr(A))\cd \tr(A)$. We can multiply the latter  by $(1-\tr(B))$ from the right,
to get $(1-\tr(A))\cd \tr(A)\cd (1-\tr(B))$. Now we \emph{must} \emph{use the commutator axioms} to commute the rightmost product in order to derive \eqref{eq:non-commutative_Modus_ponens}.

Indeed, given the product of two formulas $f\cd g$, it is possible to show by induction on the size of $f,g$, that using the commutator axioms one can derive with a
size $|f+g|$ non-commutative PC proof the formula $g\cd f$.  
\medskip 

\uline{\emph{Step 2}}: Here we repeat almost precisely the same idea as in Step 2 of the proof of Theorem \ref{thm:IPS-simulates-EF}. We have a tree-like non-commutative PC refutation  of $\tr(\varphi)$ (that possibly uses the commutator axioms) and we wish to turn it into a non-commutative formula that constitutes an IPS refutation  of $\tr(\varphi)$. We do this by constructing a non-commutative formula whose underlying graph is the same underlying proof-graph, as we did before.     
\qed 

\subsection{Frege Quasi-Polynomially Simulates the Non-Commutative IPS}

\begin{theorem}[Li-Tzameret-Wang \cite{LTW15-CCC}]
\label{thm:intro:Frege_sim_ncIPS}
Let $\phi$ be an unsatisfiable CNF formula and  $f_1,\ldots, f_m$ be the non-commutative formulas corresponding to its clauses via $tr(\cd)$. If there is a non-commutative IPS refutation of size $s$ of $f_1,\ldots, f_m$ over $GF(2)$,  then there is a Frege proof of size $s^{O(\log s)}$ of the tautology
$\neg\phi$. 
\end{theorem}

For low-degree non-commutative
IPS refutations, the proof of Theorem \ref{thm:intro:Frege_sim_ncIPS}
achieves in fact a slightly stronger simulation than stated.  Specifically, when the degree of the non-commutative IPS refutation is logarithmic in $s$, the Frege proof is of polynomial-size in $s$ (see \cite{LTW15-CCC}
for details).

The higher-level  argument is a  short Frege proof of the correctness of the Raz-Shpilka
\cite{RazShpilka05} deterministic PIT algorithm.
This resembles the discussion in Section \ref{sec:PIT_Bridge} about  PIT for (commutative) circuits. Indeed, the argument can be viewed as a realization---for the non-commutative case---of Grochow-Pitassi \cite{GrochowPitassi14} PIT-axioms framework (Section \ref{sec:PIT_Axioms}). The actual proof of Theorem \ref{thm:intro:Frege_sim_ncIPS} is rather technical and long because one  needs to prove properties of the Raz-Shpilka PIT algorithm for non-commutative formulas within the restrictive framework of propositional (Frege) proofs. Let us sketch the main ideas in the proof. \smallskip 

Our goal is to prove $\neg \phi$ in Frege, given a  non-commutative IPS refutation $\pi$ of $\phi$. The proof uses the following   five steps. \uline{First}, we \emph{balance} the non-commutative IPS $\pi$, so that its depth is logarithmic in its size. This follows more or less   Hrube\v s and Wigderson's \cite{HW14}
construction. 
%
\uline{Second},  consider the  balanced $\pi$, which is a non-commutative polynomial identity over $GF(2)$, as a \emph{boolean tautology}, by replacing plus gates with XORs and product gates with ANDs. 
\uline{Third}, we use the  so-called \emph{reflection principle} to reduce the task of efficiently proving $\neg\phi$ in Frege to the following task: show that any non-commutative formula identity over $GF(2)$, considered as a boolean tautology, has a short Frege proof (this part was discussed---for the commutative case---in Section \ref{sec:PIT_Bridge}). \uline{Fourth}, for technical reasons we turn our non-commutative polynomial identities over $GF(2)$ (considered as boolean tautological formulas) into a sum of \emph{homogenous} identities. This is the \emph{only} step that is responsible for the \textit{quasi-polynomial size increase} in Theorem \ref{thm:intro:Frege_sim_ncIPS}.
For this step we use an efficient Frege simulation of a result by Raz \cite{Raz13-tensor} who showed how to transform an algebraic
 formula into (a sum of) homogenous formulas in an efficient manner.

The \uline{fifth} and final step  is to actually construct
 short Frege proofs for homogenous non-commutative identities over $GF(2)$ translated into propositional tautologies.  
To this end we construct an efficient  Frege proof of the correctness of the Raz-Shpilka PIT algorithm for non-commutative formulas  \cite{RS04}.

\medskip 

In conclusion, the fact that  IPS written as a non-commutative formula (with the additional commutator
axioms) characterizes Frege proofs, naturally motivates studying $\cC$-IPS for various restricted classes
\mar{C-IPS undefined} $\cC$ of algebraic circuits. Lower bounds for such proofs intuitively correspond to lower bounds for restrictions of the Extended Frege proof system. This
is the content of the next section. 

\section{Lower Bounds on Fragments of IPS}\label{sec:lower bounds}
In Section \ref{sec:VNP} we have seen that proving super-polynomial lower bounds on the size of IPS certificates (written as algebraic
circuits) would imply a separation of \VP\ from \VNP. On the other hand, in Section \ref{sec:noncommIPS}
we have seen that already proving lower bounds on  IPS certificates when they are written as non-commutative
formulas and augmented with the commutator axioms would imply Frege lower bounds. It is then natural to attempt to obtain lower
bounds on IPS refutations where the certificates are written as an algebraic circuit from a restricted circuit class $\cC$. Recall the notation $\cC$-IPS from Definition \ref{def:orig-IPS},
denoting that the IPS certificate  $C(\vx,\vy,\vz)$ is taken
from the class $\cC$.\footnote{Note that there is a slight technical
difference between requiring that $C(\vx,\vy,\vz)$ is taken from
$\cC$ and requiring that $C(\vx,\vF(\vx),x_1^2-x_1,\ldots,x_n^2-x_n)$
is taken from $\cC$. In $\cC$-IPS we require the former.} If the ``placeholder'' variables $\overline y,\overline z$ in $C$ are linear we call the certificate a $\cC$-\lIPS\ certificate.
 Super-polynomial lower bounds on the size of $\cC$-\lIPS\
refutations were recently shown by Forbes, Shpilka, Tzameret and Wigderson \cite{FSTW16} when $\cC$ is the class of read once (oblivious) algebraic branching programs (roABPs), multilinear formulas and diagonal circuits. We now survey some of these lower bounds.

Let us describe the main strategy behind the proofs, which is new, and exemplifies the potential of the algebraic complexity-based approach  in proof complexity. One feature of these proof-size lower bounds is that they stem almost directly from  circuit-size lower bounds. 

Assume that $f(\overline
x)=0$ has no 0-1 solutions over some field $\F$, and let
\begin{equation}\label{eq:ns-lower bound}
    g(\overline x)\cd f(\overline x)+\sum_{i=1}^n h_i(\overline x)\cd
    (x_i^2-x_i) = 1\,,
\end{equation}
be the Nullstellensatz (equivalently, \lIPS) refutation of $f(\overline
x)=0$. We are going to lower bound the size of circuits computing $g(\overline x)$. If we restrict our
attention in  \eqref{eq:ns-lower bound} to only 0-1 assignments, then the boolean axioms vanish, and we have 
\begin{equation}\label{eq:NS-0-1}
    g(\overline x)\cd f(\overline x) = 1\,,~~~~~\hbox{for } \overline x\in\zo^n,
\end{equation} 
where \eqref{eq:NS-0-1} is now a \emph{functional identity} (in contrast to a \emph{formal} identity between polynomials). That is, we consider $g(\overline x)\cd f(\overline x)$ as a function from $\zo^n$
to $\F$, and conclude that this is the constant 1 function. Hence
\begin{equation*}\label{eq:one-over-f}
    g(\overline x) = \frac{1}{f(\overline x)}\,,~~~~~\hbox{for } \overline x\in\zo^n\,.
\end{equation*}
Therefore, to lower bound the algebraic circuit size of  Nullstellensatz refutations of $f(\overline x)$
it suffices to lower bound the algebraic circuit size of every polynomial that computes the \emph{function}
$1/f(\overline x)$ over 0-1 assignments. Since we wish to prove a lower bound for a family of polynomials computing a certain
function over 0-1, instead of a lower bound for a specific formal polynomial, this kind of lower bound is called a \emph{functional lower bound} (see also \cite{GR00,FKS15}). Forbes et al.~\cite{FSTW16} showed that for $f(\overline x)$ being a variant of the subset-sum principle $\sum_{i=1}^n \alpha_i x_i -m$, for $\alpha_i\in\F$ and  $m\not\in\set{\sum_{i=1}^n \alpha_i x_i\;:\;\overline x\in\zo^n}$, one can obtain strong functional lower bounds on the algebraic circuit-size of $1/f(\overline x)$, for certain circuit classes, and thus concluded IPS refutation lower bounds for these circuit classes.

The lower bounds obtained in \cite{FSTW16}, stated below, are for IPS over multilinear formulas and read once (oblivious) algebraic branching programs (roABP). A \emph{multilinear formula} is simply an algebraic formula (Section \ref{sec:algebraic_circuits}) in which every node computes a multilinear polynomial. For the definition of roABPs and the proof of the lower bounds we refer the reader to \cite{FSTW16}.



\begin{theorem}[Forbes-Shpilka-Tzameret-Wigderson \cite{FSTW16}]\label{thm:FSTW16}
Let $n\ge 1$ and $\F$ be a field with characteristic
bigger than $2n \choose n$. Suppose that   $f(\overline x,\overline z) =  \sum_{i<j\in[2n]}z_{i,j}x_ix_j-m$ is a polynomial over $\F$ that has no 0-1 roots. Then, any $\cC$-\lIPS\ refutation of $f(\vx,\vz)$ requires:\vspace{-6pt}
\begin{enumerate}
\item $n^{\Omega(\log n)}$-size when $\cC$ is the class
of  multilinear formulas; 

\item $2^{n^{\Omega(1)}}$-size when $\cC$ is the class of constant-depth multilinear formulas; and 

\item $2^{\Omega(n)}$-size when $\cC$ is the class of roABPs (in every variable order). 
\end{enumerate}
\end{theorem}

Forbes et al.~\cite{FSTW16} also obtained nontrivial upper bounds on $\cC$-\lIPS for $\cC$ the class of multilinear formulas and roABPs. In particular they showed that both of these
$\cC$-\lIPS proof systems are strictly stronger than Nullstellensatz (measured by total number of monomials) and admit polynomial-size refutations of the subset-sum variant  $\sum_{i=1}^n \alpha_i x_i - m$, for $\alpha_i,m\in\F$ and $m\not\in\{\sum_{i=1}^n\alpha_i
x_i\;:\; \vx \in\zo^n\}$. For more details see \cite{FSTW16}.

\section{PIT and Proof Complexity}\label{sec:PIT}
We already discussed the polynomial identity testing (PIT) problem
in the context of both IPS and the non-commutative IPS. There, we
were interested in the following question:
\begin{quote}
\textit{Can propositional proofs  efficiently prove the correctness of a PIT algorithm
for a given circuit class?}
\end{quote} 
We have seen in Section \ref{sec:PIT_Axioms} that the PIT axioms
 capture the statements that express the correctness of  a PIT algorithm (formally, a circuit for PIT). In other words, providing short Extended Frege proofs for the PIT axioms would de facto mean that Extended Frege efficiently proves the correctness of (some) polynomial-size
circuits for PIT; from which it follows that Extended Frege polynomially
simulates IPS.
Subsequently, in Section \ref{thm:intro:Frege_sim_ncIPS}, we showed
that  Frege \emph{does }admit efficient (quasi-polynomial) proofs of the correctness of the Raz-Shpilka deterministic PIT algorithm for non-commutative formulas. This, in turn,  implies that Frege quasi-polynomially simulates the non-commutative IPS (over $GF(2)$). 

\subsection{Proof Systems for  Polynomial Identities}

Hrube\v s and Tzameret
 \cite{HT08} asked the
following question concerning the connection between proof
complexity and the PIT problem: \begin{quote}
\textit{Can we efficiently \emph{prove} polynomial identities? And
specifically, is there a sequential proof system admitting polynomial-size
proofs for all polynomial identities?}
\end{quote} 
In other words, this question asks whether the PIT problem
 admits short proofs, and specifically, whether there is a simple sequential proof system to witness that PIT has
short proofs.

We know that an efficient \emph{probabilistic} algorithm for PIT exists, due to Schwartz and Zippel \cite{Schwartz80,Zippel79}: when the field is sufficiently large, with high  probability, two different polynomials will differ on a randomly chosen field assignment. However, whether the PIT problem is in \P, namely is solvable in \emph{deterministic} polynomial-time, is of course a  major open problem in complexity and derandomization theory. In fact, it is not even known whether there are sub-exponential-size  \emph{witnesses} that two algebraic formulas compute the same polynomial; namely, whether there is a \emph{non-deterministic} sub-exponential-time algorithm for PIT, and  in yet other words, whether PIT is in  \NP\ (note that  PIT \emph{is} known to be in \coRP$\subseteq$\coNP). \mar{Should mention Kabanets-Impagliazzo and its recent improvements by Jansen-Santhanam (few years ago) and Carmosino et al (last year?) here. Also, it may be worth noting that if PIT is in NP, then PIT is in ZPP...}
   
\mar{Should have discussed PIT and BPP above, in the definitions section; or better here: namely move all PIT stuff to here.}  Hrube\v s-Tzameret's work \cite{HT08} investigated the PIT$\in?\NP$\mar{\
\mbox{} \\ \\  Better typography for $\in?\NP$} question from the proof-complexity perspective: assuming that PIT does posses short witnesses, is it the case that a proof system using only symbolic manipulation (resembling a Frege proof) is enough to provide these short witnesses? And if not, can we prove lower bounds
on such proofs? Such lower bounds would at least delineate the methods
that will work for efficiently proving polynomial identities and those that will
not.

 The work in \cite{HT08}, as well as the subsequent work \cite{HT12},  set out to define the analogs of Frege and Extended Frege for the PIT problem that we shall call \textit{PI proofs} (for Polynomial
Identity Proofs; originally these systems were called \emph{arithmetic proofs}): just as Frege and Extended Frege prove boolean tautologies
by deriving new tautological formulas (and circuits, resp.), PI proofs
prove polynomial identities by deriving new  identities
between algebraic formulas (and circuits, resp.). 

Let us first describe
 the analog of Frege for PIT, namely the PI proof operating
with algebraic formulas, denoted  $\PPF$ (where
$f$ stands for ``formulas"). $\PPF$ is a sequential proof system whose axioms are the polynomial-ring axioms and whose
derivation rules express the properties of the equality symbol. Each
proof-line in the system is an equation between two algebraic formulas
(or circuits; see below) $F,G$ computing polynomials over a given field $\F$  
written as $F=G$. The proof system is sound and complete for
true polynomial identities:
\begin{theorem}[\cite{HT08}] Let $\F$ be a field. For any pair $ F,G$ of algebraic formulas, there is a PI proof in the system  $\PPF$ of $F=G$ iff $F$ and $G$ compute
the same polynomial.
\end{theorem}
The specific description of the rules and axioms of the PI proof system $\PPF$ are quite natural. The inference rules of $\PPF$ are (with $F,G,H$ formulas; where an equation below a line can be inferred from the one above the line):
\newcommand{\mez}{\qquad}
\begin{gather*}
 \frac{F=G}{G=F} ~~~~~~~~~~~~  \frac{F=G\qquad G=H}{F=H} ~~~~~~~~~~~~~\frac{F_1=G_1 \qquad  F_2=G_2}{F_1 \circ F_2= G_1\circ G_2}\hbox{~ for $\circ\in\{+,\cd\}$}\,.
\end{gather*}
And the axioms of \PPC\ express reflexivity of equality  ($F=F$), commutativity and associativity of addition  and product  $(F\circ G=G\circ F$, and $F\circ(G\circ H) = (F\circ G)\circ H,$ for $\circ\in\{+,\cd\}$),  distributivity $(F\cdot (G+H) = F\cdot G+F\cdot H)$, zero element ($F+0 = F$, $F\cdot 0 = 0$), unit element ($F\cd 1 = F$) and true identities in the field ($a\circ b = c$, for $\circ\in\{+,\cd\}$ and $a,b,c\in\F$). 


    
A \emph{PI proof} $\PPF$ is thus a sequence of equations $F_1=G_1,\, F_2=G_2,\dots, F_k=G_k$, with $F_i, G_i$ formulas, such that every equation is either an axiom or was obtained from previous equations by one of the inference rules. The \emph{size} of a proof  is the total size of all formulas appearing in the proof. 
It is easy to see that, just like a Frege proof, a PI proof can be verified for correctness in polynomial-time (assuming the field has efficient representation; e.g., the field of rational numbers). \smallskip 

It is important to notice the distinction between PI proofs and propositional proofs: PI proofs prove polynomial identities
(a language in \coRP) while propositional proofs prove boolean tautologies (a language in \coNP). See more on this in Section \ref{sec:Arithmetic_Fragment} below.

The analog of Extended Frege for PIT, denoted \PPC, is identical to $\PPF,$ except
that it operates with equations
between algebraic \emph{circuits} instead of algebraic formulas (similar to Je\v rabek's Circuit Frege \cite{Jer04}, formally, one needs to add another rule to such a system to be able to symbolically
manipulate circuits, namely to merge two separate but identical sub-circuits into a single sub-circuit; see \cite{HT12} for the details.)

It turns out that PI proofs are in fact quite  strong. First, \cite{HT08} only demonstrated lower bounds on very restricted fragments of PI proofs, and  apparently it is quite hard to go beyond these restricted fragments of PI proof systems  (assuming
any nontrivial lower bound even exists). Furthermore, PI proofs were
found to admit short proofs for many non-trivial polynomial identities
(like identities based on symmetric polynomials). Moreover,  PI proofs are able to ``simulate'' PIT algorithms for restricted algebraic circuit
classes; specifically, Dvir and Shpilka's PIT algorithm for restricted depth-3 algebraic circuits \cite{DS04}. But more importantly, PI proofs were shown in \cite{HT12} to efficiently simulate many of the classical structural results on  algebraic circuits. 

In particular, PI proofs  $\PPC$ operating with equations between algebraic circuits, efficiently simulate the following 
constructions:
(i) \emph{homogenization }of algebraic  circuits (implicit in \cite{Str73}); (ii) Strassen's technique for \emph{eliminating division gates }over large enough fields (also in \cite{Str73}); (iii) eliminating division gates over small fields---this is done by simulating large fields in small ones; and (iv) \emph{balancing algebraic  circuits} (Valiant et al.~\cite{VSB+83}; see also \cite{Hya79}). Most notably, the latter result gives a strong \emph{depth reduction} for polynomial-size $ \PPC $ proofs to polynomial-size $ O(\log^2 n) $-depth  $ \PPC $ proofs  and a quasi-polynomial simulation of $\PPC$ by $\PPF$. This is one important point where the PI proof systems   differ from Frege and Extended Frege, for which no  such simulation is known.


Since depth reduction is the most important of these
results, let us state this more formally: 

\begin{theorem}[Depth reduction for PI proofs \cite{HT12}]\label{thm:depth-reduction} Assume that $F,G$ are circuits of (syntactic\footnote{The syntactic degree of an algebraic circuit is the maximal total degree of a monomial computed after multiplying out all brackets in the circuit (without cancelations of monomials).}) degree $\leq d$ and depth $\leq t$. If $F=G$ has a    $\PPC$ proof of size $s$ then it has  a $\PPC$ proof of size $\poly(s,d)$ and depth $O(t+\log s\cd\log d +\log^2 d) $.
\end{theorem}

Intuitively, one can think of this theorem as showing that $\VNC^2$-PI
proofs  are equal in strength to $\VP$-PI proofs, similar
to the strong depth collapse manifested for algebraic circuits by Valiant et al.~\cite{VSB+83} who showed that  $\VNC^2=\VP$ (where $\cc{VNC}^2$ is defined similar to $\VP$ except that the depth of the circuits computing $f_n$ is required
to be $O(\log^2 n)$).

As we now discuss, this also had implications for understanding \emph{propositional proofs}. 

\para{Algebraic Fragments of Propositional Proofs}\label{sec:Arithmetic_Fragment}
Recall the Frege proof system described in Section \ref{sec:proof-complexity-background}.
Each Frege proof-line is a propositional tautological formula, which
is either a (substitution instance of an) axiom \mar{substitution
instance-define} or was derived by the modus ponens  rule.
As mentioned before, Reckhow \cite{Reckhow76} proved that it does
not matter which derivation rules and axioms we use, nor even the
specific logical connectives (gates) used: as long as we use a finite number of rules and axioms\footnote{The number of axioms and rules is finite, but they obviously induce infinite many substitution instances of axioms and rules, since the axioms and rules are closed under substitution of the variables in the axioms and rules by formulas} and the rules and axioms are (implicationally) complete, every two Frege systems are polynomially equivalent. 

Now, consider the PI proof system $\PPF$, and assume the underlying field is $GF(2)$. In this case, \emph{every PI proof-line becomes a boolean tautology}, where ``$+$'' becomes the logical gate XOR, ``$\cd$'' becomes  AND  and ``$=$'' becomes the logical equivalence gate $\equiv$ (indeed, note that over $GF(2)$ the axioms become propositional
tautologies). This then means that PI proofs over $GF(2)$
\emph{are by themselves propositional Frege proofs}. The converse,
on the other
hand, is not true: not all Frege proofs are arithmetic proofs over $GF(2)$, because
PI proofs over $GF(2)$ are \emph{not} complete for the set of tautologies. For instance, $x_i^2+x_i=0$ is, over $GF(2)$ the boolean tautology $(x_i\land x_i)\oplus x_i \equiv \mathsf {false}$\, over $GF(2)$, but it is not a true identity between (formal) polynomials
and thus cannot be proved by a PI proof.
In fact, Frege system is equivalent to the PI-system $\PPF$ over
$GF(2)$ \emph{augmented
with the boolean axioms} $x_i^2+x_i$; and similarly for Extended Frege
and $\PPC$ over $GF(2)$.

Considering PI proofs as the ``algebraic fragment'' of propositional proofs gives us a new understanding of propositional proofs, and should hopefully shed more light on the complexity of Frege proofs. As mentioned above, it shows for instance a strong depth collapse, namely, that additional depth (beyond $O(\log ^2 n)$) does not help to decrease the complexity of proofs. It is specifically useful for upper bounds questions on propositional proofs: if we can efficiently prove an algebraic identity over $GF(2)$ with a PI proof we can do the same for propositional proofs. This observation was used in \cite{HT12} to give a polynomial-size and depth-$O(\log ^2 n)$ Extended Frege proof of the determinant identities $Det(A)\cd Det(B)=Det(AB)$ and other linear-algebraic statements such as the matrix inverse principle $AB=I_n \to BA=I_n$.  By, essentially, unwinding depth-$O(\log ^2 n)$ \emph{circuits} into quasi-polynomial-size \emph{formulas}, one can obtain quasi-polynomial-size Frege proofs of the same statements. These results give presumably tight upper bounds for the proof complexity of linear algebra, because, for example, Bonet, Buss and Pitassi  conjectured that Frege does not admit polynomial-size proofs of these identities  \cite{BBP95}.


\section{Conclusion and Open Problems}\label{sec:conclusion}
In this survey we demonstrated the  emerging algebraic complexity approach to proof complexity. It is natural to expect that this close interaction between algebraic and proof complexity will  continue to contribute  new insights to  proof complexity. Already now very interesting and sometimes surprising new ideas came out from this interaction. In particular, we have seen that proof complexity lower bounds (for IPS restricted subsystems) are drawn almost directly from algebraic circuit lower bounds; new connections between computation and proofs, showing that some proof complexity lower bounds (IPS) imply computational lower bounds and complexity class separations (\VP$\neq$\VNP); and conversely, proving that certain lower bounds on weak computational models (non-commutative formulas) would imply strong Frege lower bounds; characterizing the ``algebraic fragments'' of  Frege and Extended Frege systems and using structural properties of algebraic circuits yield a better understanding of the power of these systems through (apparently) tight short proofs for basic statements in linear algebra. All of these results have been achieved using methods from algebraic complexity.

The big challenge ahead is of course to find out whether the algebraic  approach can eventually lead to lower bounds on Frege and Extended Frege, or conversely help to at least establish a formal (unconditional) so-called `barrier' against proving such lower bounds (e.g., by showing that Extended Frege lower bounds imply strong explicit circuit lower bounds). But before this seemingly formidable challenge, there are many important intermediate problems which seem relatively feasible at the moment, and whose solution will advance the frontiers of our understanding. We end by listing some of these problems. \vspace{-2pt}
\begin{itemize}
\item Can we obtain size lower bounds on constant-depth Frege with modular gates proofs ($\ACZ[p]$-Frege proofs)? This problem has been open for decades, despite the known $\ACZ[p]$-circuits lower bounds. It is quite conceivable that algebraic techniques may be of help
on this (cf.~\cite{MP97,BKZ12}).  

\item 
Can we establish size lower bounds on $\cC$-IPS (linear or not) refutations for  natural encodings of CNFs, for restricted circuit classes $\cC$? The lower bounds from \cite{FSTW16} hold only for a single hard axiom, and not CNFs.  

\item 
Can we extend the $\cC$-IPS lower bounds to  ``dynamic'' versions of $\cC$-IPS? For instance, can we prove lower bounds on PC refutations operating with multilinear formulas or roABPs as in \cite{RT06,Tza11-I&C}?

\item 
Lower bounds on PI proofs of polynomial identities? Almost no lower bound is known for these ``algebraic fragments'' of Frege and Extended Frege. 

\item 
Just like PI proofs are proofs for the (algebraic) language of polynomial identities, it is very interesting to study the complexity of proof systems for other algebraic languages. Two examples of such proof systems are the proof systems for matrix identities investigated in \cite{LT13}, and the proof system for non-commutative rational identities  defined in \cite{GGOW15}. Can we prove strong proof-size lower bounds on these  systems? Can we connect these systems further to propositional proof complexity or algebraic circuit complexity?  

\end{itemize}

\section{Acknowledgements}
We thank Stephen Cook, Kaveh Ghasemloo, Amir Shpilka and Avi Wigderson for very helpful discussions and clarifications, and Michael Forbes for very useful discussions and comments  on a preliminary draft. We would also like to thank Neil Immerman for his careful reading and useful comments on this survey.  Finally we are especially grateful to Joshua Grochow for many conversations and for answering our many questions  that greatly improved this survey. 

{\footnotesize
\bibliographystyle{apalike}
\bibliography{PrfCmplx-Bakoma}

\begin{thebibliography}{}

\bibitem[Alekhnovich and Razborov, 2001]{AlekhnovichRazborov01}
Alekhnovich, M. and Razborov, A.~A. (2001).
\newblock Lower bounds for polynomial calculus: Non-binomial case.
\newblock In {\em \FOCS{2001}}, pages 190--199.

\bibitem[Beame et~al., 1996]{BeameIKPP96}
Beame, P., Impagliazzo, R., Kraj{\'{\i}}{\v{c}}ek, J., Pitassi, T., and
  Pudl{\'a}k, P. (1996).
\newblock Lower bounds on {H}ilbert's {N}ullstellensatz and propositional
  proofs.
\newblock {\em Proc. London Math. Soc. (3)}, 73(1):1--26.
\newblock \pFOCS{1994}.

\bibitem[Bonet et~al., 1995]{BBP95}
Bonet, M.~L., Buss, S.~R., and Pitassi, T. (1995).
\newblock Are there hard examples for {F}rege systems?
\newblock In {\em Feasible mathematics, II (Ithaca, NY, 1992)}, volume~13 of
  {\em Progr. Comput. Sci. Appl. Logic}, pages 30--56. Birkh\"auser Boston,
  Boston, MA.

\bibitem[Buss et~al., 2001]{BussGIP01}
Buss, S.~R., Grigoriev, D., Impagliazzo, R., and Pitassi, T. (2001).
\newblock Linear gaps between degrees for the polynomial calculus modulo
  distinct primes.
\newblock {\em \JCSS}, 62(2):267--289.
\newblock \pCCC{1999}.

\bibitem[Buss et~al., 1996]{BussIKPRS96}
Buss, S.~R., Impagliazzo, R., Kraj{\'{\i}}{\v{c}}ek, J., Pudl{\'{a}}k, P.,
  Razborov, A.~A., and Sgall, J. (1996).
\newblock Proof complexity in algebraic systems and bounded depth {F}rege
  systems with modular counting.
\newblock {\em \ComputationalComplexity}, 6(3):256--298.

\bibitem[Buss et~al., 2015]{BKZ12}
Buss, S.~R., Kolodziejczyk, L.~A., and Zdanowski, K. (2015).
\newblock Collapsing modular counting in bounded arithmetic and constant depth
  propositional proofs.
\newblock {\em Transactions of the AMS}, (367):7517--7563.

\bibitem[Clegg et~al., 1996]{CleggEI96}
Clegg, M., Edmonds, J., and Impagliazzo, R. (1996).
\newblock Using the {G}roebner basis algorithm to find proofs of
  unsatisfiability.
\newblock In {\em \STOC{1996}}, pages 174--183.

\bibitem[Clote and Kranakis, 2002]{CK02}
Clote, P. and Kranakis, E. (2002).
\newblock {\em Boolean functions and computation models}.
\newblock Texts in Theoretical Computer Science. An EATCS Series.
  Springer-Verlag, Berlin.

\bibitem[Cook, 1975]{Coo75}
Cook, S.~A. (1975).
\newblock Feasibly constructive proofs and the propositional calculus
  (preliminary version).
\newblock In {\em STOC}, pages 83--97.

\bibitem[Cook and Reckhow, 1974a]{CookReckhow74a}
Cook, S.~A. and Reckhow, R.~A. (1974a).
\newblock \aaaa{a}{O}n the lengths of proofs in the propositional calculus
  (preliminary version).
\newblock In {\em \STOC{1974}}, pages 135--148.
\newblock For corrections see Cook-Reckhow~\cite{CookReckhow74b}.

\bibitem[Cook and Reckhow, 1974b]{CookReckhow74b}
Cook, S.~A. and Reckhow, R.~A. (1974b).
\newblock \aaaa{b}{C}orrections for ``{O}n the lengths of proofs in the
  propositional calculus (preliminary version)''.
\newblock {\em {SIGACT} News}, 6(3):15--22.

\bibitem[Cook and Reckhow, 1979]{CookReckhow79}
Cook, S.~A. and Reckhow, R.~A. (1979).
\newblock The relative efficiency of propositional proof systems.
\newblock {\em J. Symb. Log.}, 44(1):36--50.
\newblock This is a journal-version of Cook-Reckhow~\cite{CookReckhow74a} and
  Reckhow~\cite{Reckhow76}.

\bibitem[Dvir and Shpilka, 2006]{DS04}
Dvir, Z. and Shpilka, A. (2006).
\newblock Locally decodable codes with 2 queries and polynomial identity
  testing for depth 3 circuits.
\newblock {\em SIAM J. on Computing}, 36(5):1404--1434.

\bibitem[Forbes et~al., 2016a]{FKS15}
Forbes, M.~A., Kumar, M., and Saptharishi, R. (2016a).
\newblock Functional lower bounds for arithmetic circuits and boolean circuit
  complexity.
\newblock In {\em 31st Conference on Computational Complexity, {(CCC)}}.

\bibitem[Forbes et~al., 2016b]{FSTW16}
Forbes, M.~A., Shpilka, A., Tzameret, I., and Wigderson, A. (2016b).
\newblock Proof complexity lower bounds from algebraic circuit complexity.
\newblock In {\em 31st Conference on Computational Complexity, {CCC} 2016, May
  29 to June 1, 2016, Tokyo, Japan}, pages 32:1--32:17.

\bibitem[Garg et~al., 2015]{GGOW15}
Garg, A., Gurvits, L., Oliveira, R., and Wigderson, A. (2015).
\newblock A deterministic polynomial time algorithm for non-commutative
  rational identity testing.
\newblock {\em CoRR}, abs/1511.03730.

\bibitem[Grigoriev, 1998]{Grigoriev98}
Grigoriev, D. (1998).
\newblock {T}seitin's tautologies and lower bounds for {N}ullstellensatz
  proofs.
\newblock In {\em \FOCS{1998}}, pages 648--652.

\bibitem[Grigoriev and Hirsch, 2003]{GH03}
Grigoriev, D. and Hirsch, E.~A. (2003).
\newblock Algebraic proof systems over formulas.
\newblock {\em Theoret. Comput. Sci.}, 303(1):83--102.
\newblock Logic and complexity in computer science (Cr\'eteil, 2001).

\bibitem[Grigoriev and Razborov, 2000]{GR00}
Grigoriev, D. and Razborov, A.~A. (2000).
\newblock Exponential lower bounds for depth 3 arithmetic circuits in algebras
  of functions over finite fields.
\newblock {\em Appl. Algebra Engrg. Comm. Comput.}, 10(6):465--487.

\bibitem[Grochow and Pitassi, 2014]{GrochowPitassi14}
Grochow, J.~A. and Pitassi, T. (2014).
\newblock Circuit complexity, proof complexity, and polynomial identity
  testing.
\newblock In {\em \FOCS{2014}}, pages 110--119.
\newblock \farXiv{abs/1404.3820}.

\bibitem[Haken, 1985]{Hak85}
Haken, A. (1985).
\newblock The intractability of resolution.
\newblock {\em Theoret. Comput. Sci.}, 39(2-3):297--308.

\bibitem[Hrube{\v s} and Wigderson, 2014]{HW14}
Hrube{\v s}, P. and Wigderson, A. (2014).
\newblock Non-commutative arithmetic circuits with division.
\newblock In {\em Innovations in Theoretical Computer Science, ITCS'14,
  Princeton, NJ, USA, January 12-14, 2014}, pages 49--66.

\bibitem[Hrube\v{s} and Tzameret, 2009]{HT08}
Hrube\v{s}, P. and Tzameret, I. (2009).
\newblock The proof complexity of polynomial identities.
\newblock In {\em Proceedings of the 24th Annual {IEEE} Conference on
  Computational Complexity, {CCC} 2009, Paris, France, 15-18 July 2009}, pages
  41--51.

\bibitem[Hrube\v{s} and Tzameret, 2015]{HT12}
Hrube\v{s}, P. and Tzameret, I. (2015).
\newblock Short proofs for the determinant identities.
\newblock {\em {SIAM} J. Comput.}, 44(2):340--383.
\newblock (A preliminary version appeared in Proceedings of the 44th Annual ACM
  Symposium on the Theory of Computing (STOC)).

\bibitem[Hyafil, 1979]{Hya79}
Hyafil, L. (1979).
\newblock On the parallel evaluation of multivariate polynomials.
\newblock {\em SIAM J. Comput.}, 8(2):120--123.

\bibitem[Impagliazzo et~al., 1999]{IPS99}
Impagliazzo, R., Pudl{\'{a}}k, P., and Sgall, J. (1999).
\newblock Lower bounds for the polynomial calculus and the gr{\"{o}}bner basis
  algorithm.
\newblock {\em \ComputationalComplexity}, 8(2):127--144.

\bibitem[Je{\v{r}}{\'a}bek, 2004]{Jer04}
Je{\v{r}}{\'a}bek, E. (2004).
\newblock Dual weak pigeonhole principle, {B}oolean complexity, and
  derandomization.
\newblock {\em Ann. Pure Appl. Logic}, 129(1-3):1--37.

\bibitem[Kraj{\'{\i}}{\v{c}}ek, 1995]{Kra95}
Kraj{\'{\i}}{\v{c}}ek, J. (1995).
\newblock {\em Bounded arithmetic, propositional logic, and complexity theory},
  volume~60 of {\em Encyclopedia of Mathematics and its Applications}.
\newblock Cambridge University Press, Cambridge.

\bibitem[Kraj\'{i}\v{c}ek, 2011]{Kra:book11}
Kraj\'{i}\v{c}ek, J. (2011).
\newblock {\em Forcing with random variables and proof complexity}.
\newblock London Mathematical Society Lecture Note Series, No.382. Cambridge
  University Press.

\bibitem[Li and Tzameret, 2013]{LT13}
Li, F. and Tzameret, I. (2013).
\newblock Generating matrix identities and proof complexity.
\newblock {\em Electronic Colloquium on Computational Complexity, TR13-185}.
\newblock arXiv:1312.6242 [cs.CC] \url{http://arxiv.org/abs/1312.6242}.

\bibitem[Li et~al., 2015]{LTW15-CCC}
Li, F., Tzameret, I., and Wang, Z. (2015).
\newblock Non-commutative formulas and frege lower bounds: a new
  characterization of propositional proofs.
\newblock In {\em 30th Conference on Computational Complexity, {CCC} 2015, June
  17-19, 2015, Portland, Oregon, {USA}}, pages 412--432.
\newblock Full Version: \url{http://arxiv.org/abs/1412.8746}.

\bibitem[Maciel and Pitassi, 1997]{MP97}
Maciel, A. and Pitassi, T. (1997).
\newblock On {${\rm ACC}\sp 0[p\sp k]$} {F}rege proofs.
\newblock In {\em Proceedings of the Annual ACM Symposium on the Theory of
  Computing 1997 (El Paso, TX)}, pages 720--729 (electronic). ACM, New York.

\bibitem[Nisan, 1991]{Nisan91}
Nisan, N. (1991).
\newblock Lower bounds for non-commutative computation.
\newblock In {\em \STOC{1991}}, pages 410--418.

\bibitem[Nordstr\"{o}m, 2015]{Nor15-siglog}
Nordstr\"{o}m, J. (2015).
\newblock On the interplay between proof complexity and sat solving.
\newblock {\em ACM SIGLOG News}, 2(3):19--44.

\bibitem[Pitassi, 1997]{Pit97}
Pitassi, T. (1997).
\newblock Algebraic propositional proof systems.
\newblock In {\em Descriptive complexity and finite models (Princeton, NJ,
  1996)}, volume~31 of {\em DIMACS Ser. Discrete Math. Theoret. Comput. Sci.},
  pages 215--244. Amer. Math. Soc., Providence, RI.

\bibitem[Raz, 2013]{Raz13-tensor}
Raz, R. (2013).
\newblock Tensor-rank and lower bounds for arithmetic formulas.
\newblock {\em J. {ACM}}, 60(6):40.

\bibitem[Raz and Shpilka, 2005a]{RazShpilka05}
Raz, R. and Shpilka, A. (2005a).
\newblock Deterministic polynomial identity testing in non-commutative models.
\newblock {\em Comput. Complex.}, 14(1):1--19.
\newblock \pCCC{2004}.

\bibitem[Raz and Shpilka, 2005b]{RS04}
Raz, R. and Shpilka, A. (2005b).
\newblock Deterministic polynomial identity testing in non commutative models.
\newblock {\em Computational Complexity}, 14(1):1--19.

\bibitem[Raz and Tzameret, 2008a]{RT07}
Raz, R. and Tzameret, I. (2008a).
\newblock Resolution over linear equations and multilinear proofs.
\newblock {\em Ann. Pure Appl. Logic}, 155(3):194--224.

\bibitem[Raz and Tzameret, 2008b]{RT06}
Raz, R. and Tzameret, I. (2008b).
\newblock The strength of multilinear proofs.
\newblock {\em Computational Complexity}, 17(3):407--457.

\bibitem[Razborov, 1998]{Razborov98}
Razborov, A.~A. (1998).
\newblock Lower bounds for the polynomial calculus.
\newblock {\em \ComputationalComplexity}, 7(4):291--324.

\bibitem[Razborov, 2015]{Razb15-annals}
Razborov, A.~A. (2015).
\newblock Pseudorandom generators hard for $k$-\textsc{DNF} resolution and
  polynomial calculus resolution.
\newblock {\em Annals of Mathematics}, 181:415--472.

\bibitem[Reckhow, 1976]{Reckhow76}
Reckhow, R.~A. (1976).
\newblock {\em On the lengths of proofs in the propositional calculus}.
\newblock PhD thesis, University of Toronto.

\bibitem[Schwartz, 1980]{Schwartz80}
Schwartz, J.~T. (1980).
\newblock Fast probabilistic algorithms for verification of polynomial
  identities.
\newblock {\em J. ACM}, 27(4):701--717.
\newblock \pEUROSAM{1979}.

\bibitem[Shpilka and Yehudayoff, 2010]{SY10}
Shpilka, A. and Yehudayoff, A. (2010).
\newblock Arithmetic circuits: A survey of recent results and open questions.
\newblock {\em Foundations and Trends in Theoretical Computer Science},
  5(3-4):207--388.

\bibitem[Strassen, 1973]{Str73}
Strassen, V. (1973).
\newblock Vermeidung von divisionen.
\newblock {\em J. Reine Angew. Math.}, 264:182--202.
\newblock (in German).

\bibitem[Tzameret, 2011]{Tza11-I&C}
Tzameret, I. (2011).
\newblock Algebraic proofs over noncommutative formulas.
\newblock {\em Inf. Comput.}, 209(10):1269--1292.

\bibitem[Valiant, 1979a]{Val79:ComplClass}
Valiant, L.~G. (1979a).
\newblock Completeness classes in algebra.
\newblock In {\em Proceedings of the 11th Annual ACM Symposium on the Theory of
  Computing}, pages 249--261. ACM.

\bibitem[Valiant, 1979b]{Val79-permanent}
Valiant, L.~G. (1979b).
\newblock The complexity of computing the permanent.
\newblock {\em Theor. Comput. Sci.}, 8:189--201.

\bibitem[Valiant, 1982]{Val82}
Valiant, L.~G. (1982).
\newblock Reducibility by algebraic projections.
\newblock {\em Logic and Algorithmic: International Symposium in honour of
  Ernst Specker}, 30:365--380.

\bibitem[Valiant et~al., 1983]{VSB+83}
Valiant, L.~G., Skyum, S., Berkowitz, S., and Rackoff, C. (1983).
\newblock Fast parallel computation of polynomials using few processors.
\newblock {\em SIAM J. Comput.}, 12(4):641--644.

\bibitem[Zippel, 1979]{Zippel79}
Zippel, R. (1979).
\newblock Probabilistic algorithms for sparse polynomials.
\newblock In {\em \EUROSAM{1979}}, pages 216--226. Springer-Verlag.

\end{thebibliography}
}

\end{document}